\documentclass[11pt]{iopart}
\usepackage{color}
\usepackage{fancyhdr}

\usepackage[]{graphicx}
\usepackage{amsfonts}

\usepackage{color}

\usepackage{graphicx}
\usepackage{makeidx}
\usepackage{indentfirst}

\newcommand{\be}{\begin{equation}}
\newcommand{\ee}{\end{equation}}
\newcommand{\en}{\end{equation}}

\newcommand{\ba}{\begin{array}}
\newcommand{\ea}{\end{array}}

\newcommand{\bea}{\begin{eqnarray}}
\newcommand{\ena}{\end{eqnarray}}

\newcommand{\beano}{\begin{eqnarray*}}
\newcommand{\enano}{\end{eqnarray*}}

\newcommand{\bei}{\begin{itemize}}
\newcommand{\eni}{\end{itemize}}

\newcommand{\bee}{\begin{enumerate}}
\newcommand{\ene}{\end{enumerate}}

\newtheorem{theorem}{Theorem}[section]
\newtheorem{coroll}[theorem]{Corollaire}
\newtheorem{lemma}[theorem]{Lemma}
\newtheorem{prop}[theorem]{Proposition}
\newtheorem{fit}[theorem]{Definition}

\newcommand{\betheo}{\begin{theorem}}
\newcommand{\entheo}{\end{theorem}}

\newcommand{\becor}{\begin{coroll}}
\newcommand{\encor}{\end{coroll}}

\newcommand{\belem}{\begin{lemma}}
\newcommand{\enlem}{\end{lemma}}

\newcommand{\befit}{\begin{fit}}
\newcommand{\enfit}{\end{fit}}

\newcommand{\beprop}{\begin{prop}}
\newcommand{\enprop}{\end{prop}}

\def \k0 {\frac{1 }{4 \pi \epsilon_0}}

\def \di { \int \int}

\def \bc {\begin{center}}
\def \ec {\end{center}}

\def\half{\mbox{\small{$\frac{1}{2}$}}}

\def \pt {\ .}

\def \dis {\displaystyle}

\def \id {\hbox{\bf \large 1} \!\! \hbox{\bf \large \sf I}}
\def \IN {\rm I \! N}

\def \IC {{\sf I} \!\!\! {\rm C}}

\def  \ni {\noindent}

\def  \lv {\left|}
\def  \rv {\right|}

\def  \li {\left<}
\def  \rs {\right>}

\begin{document}

\title[Photon-added  coherent states]
{Generalized photon-added associated  hypergeometric coherent states: characterization and relevant properties}
  
\author{
 K Sodoga \footnote{ksodoga@tg.refer.org}$^{a,b}$,
 I. Aremua$^{a,b}$ \footnote{claudisak@yahoo.fr} and
 M N Hounkonnou$^b$ \footnote[1]{To
whom correspondence should be addressed: norbert.hounkonnou@cipma.uac.bj, with copy to hounkonnou@yahoo.fr}
   }

\address{
$^a$ Universit\'e de Lom\'e Facult\'e des Sciences, D\'epartement de Physique,\\
 Laboratoire de Physique des Mat\'eriaux et de M\'ecanique Appliqu\'ee,\\
 02 BP 1515 Lom\'e, Togo\\\quad \\
$^b$ International Chair in Mathematical Physics   and Applications
(ICMPA--UNESCO Chair), University of Abomey-Calavi, 072 B.P. 050  Cotonou, Benin }

\begin{abstract}
This paper presents  the construction of a new set of   generalized photon-added coherent states related to
associated hypergeometric  functions 
introduced in our previous work (Hounkonnou M N and  Sodoga K \, 2005 \, {\it J. Phys. A: Math. Gen}\,  {\bf 38}\,  7851).
These states satisfy all required mathematical and physical properties. The associated Stieltjes power-moment problem is explicitly solved by 
using Meijer's G-function 
and the Mellin inversion theorem. 
Relevant quantum optical  and thermal characteristics   are investigated. 
The formalism is applied to  particular cases of the associated  Hermite, Laguerre,  Jacobi polynomials and
hypergeometric functions. Their corresponding states exhibit
sub-Poissonian  photon number statistics.
\end{abstract}

\date{\today}
\maketitle

\section{Introduction}

\ni Coherent states (CS)  attracted much attention in the recent decades since their introduction by  
Schr\"odinger in 1926 \cite{Schrodinger} in the context of the one-dimensional harmonic oscilator, as the Heisenberg minimum-uncertainty 
states. 
  Later they were
rediscovered by Klauder, Glauber, and Sudarshan at the beginning of the 1960s
\cite{Glauber}-\cite{Suda}.
Since then, CS
and their various generalizations \cite{Barut}-\cite{Vourdas}  are spread in the literature  
on quantum physics,  atomic  and condensed
matter physics \cite{Klauder_App, Gazeau}, as well as in   quantization   problems \cite{Berezin}-\cite{Aremua2} (and references listed therein) with related mathematical tools \cite{Ali}. For more details on the CS construction and their various generalizations, see also  \cite{Aragone, Nieto}.
CS  defines  an overcomplete family of vectors in the   Hilbert space describing a physical problem. The canonical CS  can be generalized as follows. Let $\{\phi_m\}^{\infty}_{m= 0}$ be an orthonormal basis of a separable Hilbert 
space $\mathfrak H,$  and $\mathfrak{D}$  an open subset of  $\,\IC.$ For $z \in \mathfrak D, $ with $z =  r e^{i \theta}, $ the 
CS are defined as 
 \bea \label{intro1}
|z\rangle = [{\cal N}(|z|^2)]^{-\half}\sum_{m = 0}^\infty {z^m \over \sqrt{\rho(m)}}\phi_m \, \in \mathfrak H
\ena
 where $\{\rho(m)\}$ is a positive sequence of real numbers and ${\cal N}(|z|^2)$ is the normalization constant.
As stated by Klauder  \cite{Klau, Klauder_App}, the  states $|z\rangle$, in order to be considered  as CS,  need to satisfy  the  following  set of minimal criteria:  
\bei
\item[(a)] Normalizability, i. e.,  $\langle z|z \rangle  = 1$
\item[(b)] Continuity in label $z$, i. e.,  $|z - z'|\to 0 \Longrightarrow |||z\rangle - |z'\rangle || \to 0$
\item[(c)] Completeness, i. e., there exists a positive weight fonction $\omega(|z|^2) $ such that 
\bea \label{intro2}
\di_{\IC} d^2 z |z \rangle \omega(|z|^2)\langle z| = \id.
\ena
\eni

Cotfas, in a recent work \cite{Cotfas},   provided a   factorization method  of associated
  hypergeometric  operators, and deduced the associated algebra and   corresponding  CS. The latters
 are  eigenstates of the annihilation  operator 
denoted $a_m$. Following Aleixo \etal \cite{Aleixo}, we  introduced, in a previous paper
 \cite{Sodoga}, a right inverse operator $a_m^{-1}$ allowing the  definition of  
generalized associated hypergeometric  CS (GAH-CS).

\ni In this paper, we present the construction of   a new  family of CS, hereafter called   generalized photon-added associated  hypergeometric CS (GPAH-CS),  obtained by adding 
photons to the conventional associated  hypergeometric CS. 
Since the work by   Agarwal and Tara \cite{Aga},   the photon-added CS (PACS) are intensively studied. 
They  are intermediate between a single-photon Fock state $|n\rangle$ and a coherent one $|z \rangle$. PACS  
have various applications in quantum optics, quantum information and computation. 
See   \cite{Dodonov}-\cite{Daoud} and references therein.      They also can generate the entangled states \cite{Berrada}.  

In some previous works \cite{Popov,popov01},  the 
relevant statistical properties and thermal expectation values were investigated and analyzed in 
the photon-added Barut-Girardello
CS, the PACS   and the generalized
 hypergeometric thermal  
 CS representations. The two latters were obtained from the definition of Appl and Schiller \cite{Appl},  
  for the pseudoharmonic oscillator and the Morse one-dimensional Hamiltonian, respectively.
   In one of our previous papers \cite{hounk-ngompe},
a family of photon added as well 
  as photon depleted CS related to the inverse of ladder operators acting on hypergeometric CS \cite{Appl}, was introduced. 
  Their squeezing and antibunching 
  properties were investigated in both standard (nondeformed) and deformed quantum optics. 
   Recently,  a new class of generalized  PACS 
was constructed by excitations on a  family of generalized CS. Their 
  non-classical features and their quantum statistical
properties  were  compared with the results obtained by Agarwalʼs PACS \cite{mojaverietal}. 

 The paper is organized as follows: first, in Section 2,  we give a brief review  on  the Cotfa's factorization method  of the  hypergeometric  operators, the algebra generated by the corresponding lowering and raising operators,  
and the   construction of the GAH-CS.
   Then, in Section 3,  we generate the GPAH-CS   by successive applications of the raising operator
on the conventional GAH-CS.  The  inner product  of 
two different GPAH-CS   is nonzero,
highlighting  that the obtained states are not mutually orthogonal. 
The associated Stieltjes power-moment problem is explicitly solved by using Meijer's G-function and the Mellin inversion theorem. 
The reproducing kernels in these  GPAH-CS are provided. 
 Some thermal statistical properties as well as 
 the photon number statistics of the GPAH-CS in terms of the  Mandel Q-parameter and the second-order correlation function, are  computed  and discussed.
Next,  in Section 4, the case of associated Hermite, Laguerre, Jacobi polynomials and hypergeometric functions is considered. 
Finally, we conclude in  Section 5 with a summary of our main results.

\section{Generalized associated  hypergeometric coherent states}
\ni We start with the following definition.
\befit
The  generalized associated  hypergeometric type CS (GAH-CS) are the CS corresponding to  the $m^{\textrm{th}} $ derivative  $\Phi_{l,m}= \kappa^m \Phi_l^{(m)}$ of the classical orthogonal polynomials $\Phi_l$ satisfying the second order differential equation of hypergeometric type:
\bea \label{eq1}
\sigma(s)\Phi''_l(s)+\tau(s)\Phi'_l(s) + \lambda_l\Phi_l(s)=0
\ena
 where $\lambda_l = -{1 \over 2}l(l-1)\sigma'' - l\tau'$, $\kappa = \sqrt{\sigma}$,  with $\sigma$ a nonnegative function;  $\sigma$ and  $\tau$  are polynomials of at most second and exactly  first degrees, respectively.
\enfit
The $\Phi_{l,m}$, called associated hypergeometric-type functions (AHF), are solutions of the eigenvalue problem $H_m \Phi_{l,m} = \lambda_l \Phi_{l,m}$ where  the 
 Hamiltonian operator $H_m$ is expressed as a second order differential operator as follows:  
\bea\label{eq2}
H_m  = -\sigma {d^2 \over ds^2} -\tau {d\over ds} + {m(m-2) \over 4}{{\sigma'}^2 \over \sigma} + {m\over 2}\tau{\sigma'\over \sigma}  - {1\over 2} m(m-2)\sigma'' - m \tau'.
\ena
The $\Phi_{l,m}$ are orthogonal 
\bea \label{eq3} 
\int_a^b \Phi_{l,m}\Phi_{k,m}\, \rho\, \rmd s
  =0,\qquad l \ne k,\qquad l, \, k \in \{ m, m+1, m+2, \ldots \}
  \ena
with respect to the positive weight function $\rho$ related to the  polynomial functions  $\sigma$ and $\tau$ by the Pearson's equation $(\sigma \rho)' = \tau
 \rho$, over  the interval $(a, b)$, which can be finite or infinite.
The  operator $H_m $  
factorizes as
   $$H_m - \lambda_m = A_m^\dag A_m,\quad
   H_{m+1}-\lambda_m = A_m A_m^\dag $$
and     fulfills the intertwining relations
   $$H_m A_m^\dag = A_m^\dag H_{m+1}\quad \textrm{ and}\quad A_m H_m = H_{m+1}A_m. $$
The mutually formal  adjoint  first-order differential operators 
 $$A_m : {\cal H}_m \longrightarrow {\cal  H}_{m+1}\quad \textrm{ and}\quad 
A_m^\dag : {\cal H}_{m+1} \longrightarrow {\cal  H}_{m}$$ 
are    defined as \cite{Cotfas}
  $$A_m = \kappa(s){\rmd \over \rmd s} - m \kappa'(s)\quad \textrm{ and}\quad
   A_m^\dag = -\kappa(s)\frac{\rmd}{\rmd s} -   {\tau(s)\over \kappa(s)}   -(m-1)\kappa'(s).$$ 
 ${\cal H}_m$ is the Hilbert space of $\{\Phi_{k,m}\}_{k \ge m},$
  for  $m \in \IN,$
  with respect to the inner product (\ref{eq3}).  We restrict ourselves to the case when   for each 
  $m\in \IN$,  ${\cal H}_m$ is dense in the  Hilbert space ${\cal H} =\{ \varphi \in  L^2(\rho(s)ds)\}$ where $L^2$ 
  is the space of square integrable functions.
The following shape invariance relations are satisfied 
 \bea\label{eq4}
 A_m A_m^\dag = A_{m+1}^\dag A_{m+1}+
r_{m+1},\quad r_{m+1} = \lambda_{m+1}-\lambda_m = -m\sigma'' -
\tau', 
\ena
 where   eigenvalues $\lambda_l$ and eigenfunctions
$\Phi_{l,m}$ are:
 \bea \label{eq5}
 \lambda_l =\sum_{k=1}^l r_k,
\quad \Phi_{l,m} = \frac{A_m^\dag}{\lambda_l - \lambda_m}\frac{A_{m+1}^\dag}{\lambda_l - \lambda_{m+1}}\cdots
\frac{A_{l-2}^\dag}{\lambda_l - \lambda_{l-2}}\frac{A_{l-1}^\dag}{\lambda_l - \lambda_{l-1}}\Phi_{l,l}
\ena
 for all $l\, \in \IN$ and $ m\,  \in
\{ 0, 1, 2, \ldots, l-1\}$, $\Phi_{l,l}$ satisfying the
relation $\displaystyle A_l \Phi_{l,l} = 0.$\\
The annihilation  and  creation operators are defined as
 \beano 
 a_m, \, a_m^\dag: {\cal H}_m \longrightarrow
{\cal H}_m, \quad a_m = U_m^\dag A_m\quad \textrm{ and}\quad a_m^\dag = A_m^\dag U_m 
\enano
within the unitary operator 
\beano
U_m: {\cal H}_m \longrightarrow {\cal H}_m,\quad U_m| l.
m \rangle = | l+1, m+1\rangle.
\enano
The states $|l, m\rangle = \displaystyle {\Phi_{l,m}\over ||\Phi_{l,m}||}$ 
are defined for all $l \ge m$ and for each $m \in \IN$.
 The mutually formal adjoint operators $a_m$ and $a_m^\dag$ 
act on the state $| l, m \rangle$ as 
\bea
 \fl a_m | l, m \rangle =
\sqrt{\lambda_l - \lambda_m}\,| l-1, m \rangle\quad \textrm{
and}\quad a_m^\dag | l, m \rangle = \sqrt{\lambda_{l+1} -
\lambda_m}\,| l+1, m \rangle,\quad l \ge m,
 \ena
and satisfy  the commutation relations:
 \bea \label{eq6}
 [a_m , a_m^\dag] = {\cal R}_m,\quad
[a_m^\dag, {\cal R}_m] = \sigma'' a_m^\dag \quad \textrm{and}\quad
[a_m, {\cal R}_m] = -\sigma'' a_m
 \ena 
where ${\cal R}_m = -\sigma''N_m -\tau'$, $N_m: {\cal H}_m \longrightarrow {\cal
 H}_m$ is the number operator defined as $N_m \Phi_{l,m}= l
 \Phi_{l,m}.$
 Remark that, when $deg \sigma = 1$,  the algebra defined by the generators in (\ref{eq6}) is isomorphic to the Heisenberg-Weyl algebra \cite{Cotfas}.
In addition to the commutation relations (\ref{eq6}),  we have
 \bea \label{eq7}
 A_m {\cal R}_m = {\cal R}_{m+1} A_m,
\ena 
and the similarity transformation 
 \bea \label{eq8}
 U_m {\cal R}_m U_m^\dag =
 {\cal R}_{m+1}+\sigma'', \quad \textrm{for\,  all}\quad  m \, \in \IN.
\ena
Setting  for all $m\, \in\, \IN$,
$| n \rangle = | m+n, m \rangle$,  $ e_n = \lambda_{m+n}-
\lambda_m,\quad m\, \in\, \IN,$
 we obtain
 \bea  \label{eq9}
\fl a_m| n \rangle = \sqrt{e_n}| n -1 \rangle,\quad a_m^\dag| n \rangle
= \sqrt{e_{n+1}}| n+1 \rangle, \quad (H_m - \lambda_m)| n \rangle =
e_n| n \rangle.\\ \nonumber
\ena 
The CS for AHF were provided by Cotfas \cite{Cotfas} as:
\bea \label{eq10}
|z\rangle =  {\cal N}(|z|^2) \sum_{n = 0}^\infty \frac{z^n}{\sqrt{\varepsilon_n}}| n\rangle
 \qquad {\cal N}(|z|^2) = \left[\sum_{n =0}^\infty \frac{|z|^{2n}}{\varepsilon_n}\right]^{-1/2}
 \ena
 for any $z$ in the open disc ${\cal C}(O, {\cal R})$ with centre $O$ and
 radius ${\cal R} = \limsup_{n \to \infty }\sqrt[n]{\varepsilon_n} \ne 0$
 where
 $\varepsilon_n = \left\{ \ba{lcl l}  1 & & \textrm{if} & n = 0\\
 e_1 e_2 \cdots e_n & & \textrm{if} & n > 0 \ea \right.$.  These CS are eigenstates of the annihilation operator $a_m,$  i.e.,
$a_m | z\rangle = z|z\rangle$.\\\\
Introducing the right-inverse operators $A_m^{-1}$, $a_m^{-1}$,  we showed in \cite{Sodoga} that the CS (\ref{eq10}) can be rewritten as 
\bea \label{eq11}
|z\rangle =  {\cal N}(|z|^2) \sum_{n = 0}^\infty {(z a_m^{-1})}^{n}| 0\rangle,
\ena
  generalized   as 
\bea \label{eq12}
 |z; {\cal R}_m\rangle & = &   \sum_{n = 0}^\infty {(z  {\cal R}_m  a_m^{-1})}^{n}| 0\rangle =
 \sum_{n = 0}^\infty \frac{z^n}{h_n({\cal R}_m)}| n\rangle
\ena
where 
\bea \label{eq13}
 h_n({\cal R}_m) = \dis \frac{\sqrt{\varepsilon_n}}{\dis \prod_{k = 0}^{n-1}({\cal R}_m + k \sigma'')} \quad \textrm{for}\quad   n \ge 1 \quad \textrm{and} \quad  h_0({\cal R}_m) = 1.
\ena
The states (\ref{eq12}) are  eigenstates
of $a_m$, 
\bea \label{eq14}
 a_m|z; {\cal R}_m \rangle = z
({\cal R}_m- \sigma'')|z; {\cal R}_m\rangle ,
 \ena
 and  satisfy the second  order differential equation
 \bea \label{eq15}
\{a_m - z ({\cal R}_m - \sigma'') \}\frac{d}{d z}|z;
{\cal R}_m \rangle = ({\cal R}_m - \sigma'')|z; {\cal R}_m\rangle.
 \ena
Furthermore, we  generalized the CS (\ref{eq12}) as: 
\bea \label{eq16}
 |z; {\cal R}_m\rangle &= &  \sum_{n = 0}^\infty {(z  f({\cal R}_m)  a_m^{-1})}^{n}| 0\rangle =   \sum_{n=0}^\infty \frac{z^n}{h_n({\cal
 R}_m)}|n\rangle 
 \ena
for any analytical function $f,$   where
 \bea \label{eq17}
h_n({\cal R}_m) = \frac{\sqrt{\varepsilon_n}}{\displaystyle
\prod_{k=0}^{n-1} f({\cal R}_m + k \sigma'')} \quad \textrm{for}\quad   n \ge 1 \quad \textrm{and} \quad  h_0({\cal R}_m) = 1.
 \ena
The CS  (\ref{eq16}) are  eigenstates
of $a_m$, 
 \bea \label{eq18}
 a_m|z; {\cal R}_m \rangle = z
f({\cal R}_m- \sigma'')|z; {\cal R}_m\rangle ,
 \ena
 and  satisfy  the condition
 \bea \label{eq19}
\{a_m - z f({\cal R}_m - \sigma'') \}{d\over dz}|z;
{\cal R}_m \rangle = f({\cal R}_m - \sigma'')|z; {\cal R}_m\rangle.
 \ena
 Taking into account the fact that ${\cal R}_m$ is an operator which acts on
 the states
 $|n \rangle$ as
 \bea \label{eq20}
 {\cal R}_m |n\rangle = [-(m+n)\sigma'' - \tau']|n
 \rangle = r_{m+n+1}|n\rangle ,
 \ena
 we  rewrite the CS (\ref{eq16}) under the form:
  \bea \label{eq21}
  |z; m\rangle = \sum_{n=0}^\infty \frac{z^n}{h_n(m)}|n\rangle, 
\ena
\quad 
where 
\bea\label{eq21bis}
 h_n(m)  = 
{\sqrt{\varepsilon_n} \over \dis \prod_{k=0}^{n-1} f(r_{m+n+1-k})} \quad \textrm{for} \quad n\ge 1
 \quad \textrm{and} \quad h_0(m) = 1.
\ena
The properties (\ref{eq18}) and (\ref{eq19}) become, 
\bea\label{eq22}
 a_m|z; m \rangle = z f(r_{m+n+2}'')|z; m\rangle 
 \ena
 \bea \label{eq23}
\{a_m - z f(r_{m+n+2}) \}{d\over dz}|z;m \rangle = f(r_{m+n+2})|z; m\rangle,
 \ena
respectively.
We  proved in \cite{Sodoga} that  the generalized coherent states (\ref{eq21}) verify the   properties of  label  continuity, 
overcompleteness,   temporal stability and   action identity.
\section{Construction of generalized photon-added associated  hypergeometric  coherent states}
\ni In this section, we construct the  generalized photon-added associated  hypergeometric type coherent states
(GPAH-CS) by repeating the action of  the raising operator $a_m^+$ on the GAH-CS. \\\\
 Let  ${\mathfrak H}_{m,p} $  be the  Hilbert subspace of ${\cal H}_m$ defined as:
\bea \label{Hilbert1}
 {\mathfrak H}_{m,p} := span\left\{\lv{n+p}\rs\right\}_{n,p \ge 0}.
\ena
\ni  Let us note that the first
$p$ number  states  $|n\rangle, n  = 0, 1, . . . , p -  1$, are missing from the state $|z,m\rangle_p \in {\mathfrak H}_{m,p}$ defined below.
 Then, the unity operator in this   subspace is to be written as \cite{Penson}
\bea \label{Hilbert2}
\sum_{n = 0}^\infty \lv{n+p} \rs \li {n + p}\rv  = {\id}_{{\mathfrak H}_{m,p}}
\ena 
where ${\id}_{{\mathfrak H}_{m,p}}$ is the identity operator on ${\mathfrak H}_{m,p}$.
Recall that  $\id_{{\mathfrak H}_{m,p}}$ is only required to be a bounded  positive operator with a densely defined inverse \cite{Ali95}.\\\\
The  GPAH-CS, denoted  by $|z,m\rangle_p,$ are defined as :
\bea \label{eq24}
|z,m\rangle_p \equiv {(a_m^+)}^p |z,m\rangle.
\ena
where $p$ is a positive integer standing for  the number of added quanta (or photons).
From (\ref{eq6}), we obtain 
\bea \label{eq25}
a_m^+\ {\cal R}_m = ({\cal R}_m + \sigma'')\ a_m^+, \quad 
a_m^+ \ f({\cal R}_m) = f({\cal R}_m + \sigma'')\ a_m^+,
\ena
for any analytical function $f$. \\ \\
The GAH-CS (\ref{eq16}) can be written as 
\beano
|z,{\cal R}_m\rangle = |0\rangle + \sum_{n = 1}^\infty  {z^n \over \sqrt{\varepsilon_n}}
\left(\prod_{k=0}^{n-1} f({\cal R}_m +k \sigma'')\right)|n\rangle,
\enano 
and using Eqs (\ref{eq9}) and (\ref{eq25}) we obtain 
\bea \label{eq26}
a_m^+ |z,{\cal R}_m\rangle =\sqrt{e_1}|1\rangle + \sum_{n = 1}^\infty z^n {\sqrt{e_{n+1}}\over \sqrt{\varepsilon_n}}
\left(\prod_{k=1}^n f({\cal R}_m +k \sigma'')\right)|n+1\rangle.
\ena
Now, applying $p$-times the operator $a_m^+$ on the state $|z,{\cal R}_m\rangle$ leads to 
\bea \label{eq27}
\fl {(a_m^+)}^p |z,{\cal R}_m\rangle = \sqrt{\varepsilon_p}\,|p\rangle + \sum_{n = 1}^\infty z^n {\sqrt{e_{n+1}.e_{n+2}\ldots e_{n+p}}\over \sqrt{\varepsilon_n}}
\left(\prod_{k=p}^{n+p-1} f({\cal R}_m +k \sigma'')\right)|n+p\rangle.
\ena
The GPAH-CS can be rewritten in the  form 
\bea \label{eq28}
|z,{\cal R}_m\rangle_p =  \sum_{n=0}^\infty {z^n \over K_n^p({\cal R}_m)}|n+p\rangle
\ena
where  the expansion coefficient is given by
\bea \label{eq29}
\fl K_n^p({\cal R}_m) = {\varepsilon_n \over \sqrt{\varepsilon_{n+p}}}{1\over \dis \prod_{k=p}^{n+p-1} f({\cal R}_m +k \sigma'')}, \quad \textrm{for}\quad  n \ge 1\quad  \textrm{and} \quad  K_0^p({\cal R}_m) = {1\over \sqrt{\varepsilon_p}}.
\ena 
Taking into account Eq. (\ref{eq20}), the GPAH-CS take the final form
\bea \label{eq30}
|z,m\rangle_p = {\cal N}_p(|z|^2,m) \sum_{n=0}^\infty {z^n \over K_n^p(m)}|n+p\rangle
\ena
where ${\cal N}_p(|z|^2,m)$ is the normalization constant  evaluated below while the expansion coefficient is given by
\bea\label{eq31}
 K_n^p(m)  = {\varepsilon_n \over \sqrt{\varepsilon_{n+p}}}
{1 \over \dis \prod_{k=0}^{n-1} f(r_{m+n+1-k})} \quad \textrm{for} \quad n\ge 1
 \quad \textrm{and} \quad K_0^p(m) = {1\over \sqrt{\varepsilon_p}}.
\ena
In the particular  case where $\deg \sigma = 1$, $ r_{m+n+1-k} = - \tau' = \textrm{constant}=c.$  
 Then, the expansion coefficient takes the form 
\bea \label{eq31bis}
\fl K_n^p({\cal R}_m) = {\varepsilon_n \over \sqrt{\varepsilon_{n+p}}}\, {1\over \dis [f(c)]^n}
 \quad \textrm{for}\quad  n \ge 1\quad  \textrm{and} \quad  K_0^p({\cal R}_m) = {1\over \sqrt{\varepsilon_p}}.
\ena 
 Let us examine the realization of the  properties  of normalization, label continuity and  overcompleteness for the GPAH-CS (\ref{eq30}) in the
  following subsections.
\subsection{Normalization and non-orthogonality}
\ni The normalization of the GPAH-CS (\ref{eq30}) is obtained by requiring  ${}_p\langle z, m|z,m\rangle_p = 1$.\\
 A straightforward computation gives the normalization constant ${\cal N}_p(|z|^2; m)$  as:
\bea \label{eq32}
{\cal N}_p(|z|^2; m) = \left[ \sum_{n = 0}^\infty {|z|^{2n} \over |K_n^p(m)|^2 }\right]^{-1/2} \pt
\ena
The inner product of two different states  $|z; m\rangle_p$ and $|z'; m\rangle_{p'}$ reads as:
\bea \label{eq33}
\fl {}_{p'}\li z';m\right.\lv z; m\rs_p & = &  {\cal N}_{p'}(|z'|^2; m) {\cal N}_p(|z|^2; m) 
\sum_{n,n' = 0}^\infty { {{z'}^\star}^{n'} z^n  \over {K_{n'}^{p'}}^\star(m) K_n^p(m)} 
\li n'+p' \right. \lv n+p \rs.
\ena 
Due to the orthonormality of the states $|n\rangle$, the inner product (\ref{eq33}) gives 
\bea \label{eq34}
\fl {}_{_{p'}}\li z';m\right.\lv z; m\rs_p & = &  {\cal N}_{p'}(|z'|^2; m) {\cal N}_p(|z|^2; m) {z'^\star}^{(p - p')}
\sum_{n = 0}^\infty {({z'}^\star z)^n \over  {K_{n +p-p'}^{p'}}^\star(m)\ K_n^p(m)}
\ena
proving that the GPAH-CS are not mutually orthogonal.
\subsection{Label continuity}
\ni  In the Hilbert  space ${\cal H}_m$, the GPAH-CS $|z,m\rangle_p$ are labeled by $p$ and $z$. The label continuity condition can then be stated as:
{\footnotesize
\bea
\fl |z -z'| \to 0 \ \textrm{and} \  |p - p'|\to 0 \Longrightarrow  
|||z,m\rangle_p - |z',m\rangle_{p'}||^2 = 2\left[1 - {\cal R}e \left({}_{_{p'}}\li z';m\right.\lv z; m\rs_p\right)\right]  \to 0.
\ena}
This is satisfied by the states  $|z,m\rangle_p$, since from Eqs. (\ref{eq32}, \ref{eq34}), we see that
\bea
p \to p' \quad  \textrm{and} \quad  z \to z' \Longrightarrow {}_{_{p'}}\li z';m\right.\lv z; m\rs_p \to 1.
\ena
Therefore the GPAH-CS  $|z,m\rangle_p$ are continuous in their labels.

\subsection{Overcompleteness}
\ni We have to determine a non-negative weight function $\omega_p(|z|^2; m)$ such that the overcompleteness  or the resolution of identity 
\bea \label{eq35}
\int_{{\IC}}\  d^2z\  \lv z; m\rs_p \omega_p(|z|^2;m) {}_{p}\li z;m\rv  = \id_{{\mathfrak H}_{m,p}} 
\equiv \sum_{n = 0}^\infty \lv n + p \rs \li n + p \rv
\ena
holds. \\\\
Taking to account the definition (\ref{eq30}) of the GPACS, we have: 
\bea \label{eq36}
\fl \int_{{\IC}}\  d^2z\  {\cal N}^2_p(|z|^2; m) \sum_{n,n' =0}^\infty 
{ {z^\star}^{n'}z^n  \over   {K_{n'}^p}^\star(m) K_n^p(m)}\ 
  \lv n' + p \rs \li n + p \rv \omega_p(|z|^2;m) = \id_{{\mathfrak H}_{m,p}} \pt
\ena
The diagonal matrix elements of the above relation, using the orthonormality of the number states  
$\lv n\rs$, gives:
\bea \label{eq37}
 \int_{\IC} \ d^2z\  {\cal N}^2_p(|z|^2; m) |z|^{2n} \omega_p(|z|^2; m) =  |K_n^p(m)|^2 \pt
\ena
This equation, after  performing the angular integration, gives:
\bea \label{eq38}
\fl \int_0^\infty dx\  x^n\  {\cal W}_p(x ; m) =  |K_n^p(m)|^2, \quad 
\quad \textrm{with} \quad 
 {\cal W}_p(x ; m) = \pi {\cal N}_p^2(x; m)\  \omega_p(x ; m)
\ena
where we use  the polar  representation $z = r e^{i\phi}$; $x$ standing  for $|z|^2 = r^2$.
The weight function $\omega_p$
 is then  related to the undetermined  moment distribution ${\cal W}_p(x ; m)$, which is the  solution of the
Stieltjes moment problem with the moments given by $ |K_n^p(m)|^2$.\\
As pointed out in \cite{6derniers}, the measure can be determined by using the Mellin transformation procedure. 
\ni Let us  rewrite (\ref{eq38}) as 
\bea \label{eq39}
\fl  \int_0^\infty  dx\  x^{n+p}\ g_p(x ; m) =  |K_n^p(m)|^2, \ \textrm{where} \  
g_p(x; m) = \pi {{\cal N}_p^2(x; m)} x^{-p} \ \omega_p(x; m)\pt
\ena
Performing the variable change  $n + p \to s -1, $ \  Eq. (\ref{eq39}) becomes:
\bea \label{eq40}
\int_0^\infty dx x^{s-1} g_p(x;m) =  |K_s^p(m)|^2 \pt
\ena
Consider the Meijer's G-function and the Mellin inversion theorem \cite{Marichev}, \cite{Prudnikov} 
{\small
\bea \label{eq41}
\fl  \int_0^\infty dx \, x^{s-1} G_{p,q}^{m,n} 
\left(\alpha x\lv \ba{c} 
a_1,... , a_n  ; a_{n+1},  ..., a_p \\
b_1,... , b_m  ; b_{m+1},  ..., b_q 
\ea \right.\right)
={1 \over \alpha^s}\ {\dis \prod_{j = 1}^m \Gamma(b_j+s) \prod_{j = 1}^n \Gamma(1 - a_j - s) \over  
\dis \prod_{j = m +1}^q \Gamma(1 - b_j -s) \prod_{j = n + 1}^p \Gamma(a_j + s)} \pt
\ena}
In the different examples  of  the next section, $ |K_s^p(m)|^2$  in the above relation can be expressed 
in terms of Gamma functions as in the second member of the Mellin inversion theorem (\ref{eq41}). Then comparing the  equations  
(\ref{eq40}) and (\ref{eq41}), $g_p(x;m)$ can be identified as the Meijer's G- function:
\bea \label{eq42}
g_p(x;m) =  G_{p,q}^{m,n} 
\left(\alpha x\lv \ba{c}  a_1,... , a_n  ; a_{n+1},  ..., a_p \\ b_1,... , b_m  ; b_{m+1},  ..., b_q  \ea \right.\right)\pt
\ena 
Once $g_p$ is determined,  the measure $\omega_p$ can be deduced as 
\bea
\omega_p(x; m) = {x^p \over \pi} {g_p(x;m) \over {\cal N}_p^2(x;m)}.
\ena
The overcompleteness of the GPAH-CS on $\mathfrak{H}_{m, p}$ displayed in (\ref{eq35}) and (\ref{eq36}) 
brings us   to discuss their  relation with 
the reproducing kernels.
\subsection{Reproducing kernels}
\ni Define the quantity $\mathcal K(z, z'):={}_{p}\li z';m\right.\lv z; m\rs_p.$ From 
\begin{eqnarray} \label{kernel0}
\fl {}_{_{p}}\li z';m\right.\lv z; m\rs_p  =   {\cal N}_{p}(|z'|^2; m)
 {\cal N}_p(|z|^2; m)\sum_{n = 0}^\infty {({z'}^\star z)^n \over  {|K_n^p(m)|^2}}
  =  \frac{{\cal N}_{p}(|z'|^2; m) {\cal N}_p(|z|^2; m)}
 { {\cal N}^2_p({z'}^\star z; m)}
\end{eqnarray}
we obtain 
\bea\label{kernel00}
\fl \overline{{}_{_{p}}\li z';m\right.\lv z; m\rs_p} = 
\frac{{\cal N}_{p}(|z'|^2;m) {\cal N}_p(|z|^2; m)}
 { {\cal N}^2_p(z^\star{z'}; m)}
:=\mathcal K(z', z).
\ena
 $\mathcal K(z, z')$ is a reproducing kernel through the following result:
  \beprop  The following properties 
  \begin{itemize}
  \item[(i)] hermiticity\quad 
$\mathcal K(z, z') = \overline{\mathcal K(z', z)}$,
   \item[(ii)]  positivity \quad 
  $\mathcal K(z, z) > 0$,
   \item[(iii)] idempotence
\bea
\int_{\IC} \ d^2z''\   \omega_p(|z''|^2; m) \mathcal K(z, z'')\mathcal K(z'', z') 
= \mathcal K(z, z')
\ena
  \end{itemize}
  are satisfied by the function $\mathcal K$ on $\mathfrak H_m.$
\enprop 
 {\bf Proof.} 
\begin{itemize}
  \item[(i)] Hermiticity:
    Using  (\ref{kernel0}) and (\ref{kernel00}), we get 
  \bea
\mathcal K(z, z') = {\mathcal K(z', z)}^{\star}.
\ena
\item[(ii)] Positivity:
From  (\ref{kernel00}), we obtain
\begin{eqnarray}
\mathcal K(z, z)= {}_{_{p}}\li z;m\right.\lv z; m\rs_p  
 =  \frac{{\cal N}_{p}(|z|^2; m) {\cal N}_p(|z|^2;m)}
 { {\cal N}^2_p(|z|^2; m)}
  =  1  > 0.
\end{eqnarray}
\item[(iii)]Idempotence:
Let ${\cal I} = \dis \int_{\IC} \ d^2z''\   \omega_p(|z''|^2; m) \mathcal K(z, z'')\mathcal K(z'', z')$. 
Then, setting $\xi_{p}(z,z'; m) = {\cal N}_p(|z|^2; m){\cal N}_p(|z'|^2; m)$ gives
\begin{eqnarray}
\fl {\cal I}&  = &  \xi_{p}(z,z'; m)  \int_{\IC} \ d^2z''\   \omega_p(|z''|^2; m)
\frac{{\cal N}^2_p(|z''|^2; m)}
{{\cal N}^2_p(zz''^{\star}; m) {\cal N}^2_p(z''z'^{\star}; m)}\cr
\fl & = &  \xi_{p}(z,z'; m)
\sum_{k,l=0}^{\infty}\int_{0}^{\infty}\int_{0}^{2\pi}\frac{e^{-i(k-l)\theta''} \, r''^{k+l}}{| K^p_k(m)|^2}
\frac{z^k(z'^{\star})^l}{| K^p_l(m)|^2}\cr
\fl & = &\xi_{p}(z,z'; m) r'' dr''\, d\theta''\omega_p(|z''|^2; m)\cr
\fl & = & \xi_{p}(z,z'; m)
\sum_{k=0}^{\infty}\frac{(\sqrt{zz'^{\star}})^{2k}}{| K^m_k(m)|^2}
\left\{\int_{0}^{\infty}\frac{x''^{k+m}}{ K^m_k(m)|^2} \, g_m(x''; m) dx''\right\}\cr
\fl & = &\frac{{\cal N}_p(|z|^2; m){\cal N}_p(|z'|^2; m)}{{\cal N}^2_p(zz'^{\star}; m)} 
 =  \mathcal K(z, z')
\end{eqnarray}
which completes the proof.
\end{itemize}
$\hfill{\square}$
\subsection{Statistical properties}
\ni  This section is devoted to the  investigation of some quantum optical features  of the GPAH-CS, 
such as the Mandel Q-parameter, 
the second order correlation function, the photon number distribution (PND), the signal-to-quantum-noise ratio (SNR). Besides, the 
relevant thermal properties are discussed by considering a quantum gas of the system for which  the density operator is elaborated.  
\subsubsection{Photon number statistics}\quad \\ 
\ni  The photon number statistics can be first studied by means of the  Mandel Q-parameter characterizing the Poisson
distribution of photons,       
defined as \cite{Mandel95}:
\bea \label{eq43}
Q = {(\Delta N_m)^2 \over \li H\rs} - 1, \quad (\Delta N_m)^2 = \langle N_m^2\rangle - \langle N_m\rangle^2.
\ena
The Mandel Q-parameter determines whether  the GPAH-CS  have a  photon number distribution. This latter is sub-Poissonian if $ -1 \le Q < 0$, 
Poissoinian if $Q = 0$, and super-Poissonian if $Q > 0$.
 \\\\
Next, from the expression   of the second order correlation function, we have 
\bea \label{eq44}
g^2 = {\langle N_m^2 \rangle - \langle N_m \rangle \over \langle N_m \rangle^2}
\ena
where the mean values are 
\bea \label{eq45}
\li N_m \rs = \, _p\li z ; m\rv N_m \lv z; m \rs_p,  \quad 
\li N_m^2 \rs =\,  _p\li z ; p\rv N_m^2 \lv z; m \rs_p .
\ena
The second order correlation function  determines either the bunching or
anti-bunching effects of the optical field: $g^2 <  1$ indicates that a photon is anti-bunching 
and thus the optical state is non-classical, while $g^2 \ge 1$ defines the classical (or random) 
optical field.
\\
One can check that for a GPAH-CS (\ref{eq30}), the expectation values of the number operator  are provided as 
\bea \label{eq46}
 \li N_m \rs& = & {\cal N}_p^2(|z|^2 ; m) \sum_{n = 0}^\infty (m+ n + p)\ {|z|^{2n}\over  |K_n^p(m)|^2 } \ , \\ \label{eq47}
\li N_m^2 \rs & = &  {\cal N}_p^2(|z|^2 ; m) \sum_{n = 0}^\infty (m+ n + p)^2\ { |z|^{2n} \over |K_n^p(m)|^2 }\pt
\ena
\ni 
Another specific key  useful for our purpose is the   PND which  represents an important tool for characterizing a given optical field 
 (see for e.g, \cite{Penson99}).   
The PND exhibiting  oscillations, which corresponds to   the probability of finding $n$ quanta in the  GPAH-CS,   
is given by the projection of 
$|z;m \rangle_p$  in the state 
$| n \rangle$ as follows \cite{Dodonov}
\bea\label{PND}
{\cal P}_n^p(x; m) = |\langle n |z;m\rangle_p|^2 = {\cal N}_p(x; m)^2\,{x^{n-p} \over |K_n^p(m)|^2}, \quad x = |z|^2.
\ena
It reduces to a Poisson distribution for the conventional CS, for $p\to 0$.\\\\
\ni Finally, we consider the SNR \cite{Dodonov, Penson99} in the GPAH-CS, which   is relevant when studying, for example, a   
nondeterministic, noiseless amplification of a CS  in the context of  use of photon addition and subtraction
as a probabilistic amplifier and it’s effects on coherent  light . When considering a CS as input and SNR as a
metric,  this latter is a remarkable tool used  to  
enhance a general signal with no added noise. It improves as well as the effect of photon addition becames important. A general construction  form for the SNR of a $p$ PACS is provided as follows \cite{bryanetal}:
\bea \label{SNR}
SNR_p+ = {\li N_m \rs - p \over \Delta N_m}
\ena
 where $p$ is  the number of added photons.\\
\ni We will see in the examples of the next section that the coefficient  $K_n^p(m)$ can be 
expressed in terms of Gamma functions.  The normalized factor, the expectation values  and the 
Mandel Q-parameter can be therefore  written in terms of  generalized hypergeometric functions $_pF_q$. 

\subsubsection{Thermal statistics} \quad \\
\ni
In quantum mechanics, the important  tool for characterizing   the probability distribution on the states 
of a physical system is
a statistical operator called density matrix, generally denoted by $\rho$. The latter is useful 
for examining the physical and chemical properties of a system (see
for e.g. \cite{Popov}, \cite{Aremua2} and references listed therein). Consider a quantum gas of the system in the thermodynamic
equilibrium with a reservoir at temperature $T$, which satisfies a quantum canonical
distribution. The corresponding normalized density operator is given as 
\bea\label{thermal00}
 \rho^{(p)} = \frac{1}{Z}\sum_{n=0}^{\infty} e^{-\beta e_n}|n + p \rangle  \langle  n + p|
\ena
where in the exponential $e_n$ is the eigen-energy, and  the partition function $Z$ is taken as the normalization constant.\\
The diagonal elements of  $\rho^{(p)}$ which are key ingredients for our purpose, 
also known as the $Q$-distribution or Husimi's distribution,  are derived  in the  GPAH-CS  basis as 
\bea\label{thermal01}
{}_{_{p}}\li z;m| \rho^{(p)} | z; m\rs_p   = \frac{{\cal N}_p^2(|z|^2 ; m) }{Z}
\sum_{n = 0}^\infty \frac{|z|^{2n} }{ |K_n^p(m)|^2 } e^{-\beta e_n}.
\ena
The normalization of the density operator leads to
\bea\label{thermal02}
\mbox{Tr} \rho^{(p)}  =  \int_{\IC}  d^2 z \, \omega_p(|z|^2; m)  \, _{p}\langle z;m| \rho^{(p)} |z;m\rangle_p =  1.
\ena
The diagonal expansion of the normalized canonical density operator  over the  GPAH-CS projector   is
\bea\label{thermal03}
 \rho^{(p)}  = \int_{\IC} d^2 z \, \omega_p(|z|^2; m) |z;m\rangle_p P(|z|^2) \, _{p}\langle z;m|
\ena
where the $P$-distribution  function  $P(|z|^2)|$ satisfying the normalization to unity condition 
\bea\label{thermal04}
 \int_{\IC} d^2 z \, \omega_p(|z|^2; m) P(|z|^2) = 1 
\ena
must  be determined.\\
Thus, given an observable $\mathcal O$, one obtains the expectation value,  i. e., the thermal average given by
\bea{\label{obaverage00}}
\langle \mathcal O \rangle_{p} = Tr ( \rho^{(p)}   \mathcal O) =  \int_{\IC} d^2 z \, \omega_p(|z|^2; m)
P(|z|^2) \,_{p}\langle z;m|\mathcal O |z;m\rangle_{p}.
\ena
Using (\ref{eq46}) and (\ref{eq47}) together, the  pseudo-thermal expectation value of the number operator $N_m,$ and of its square $N^2_m, $
given by $\langle   N_m \rangle^{(p)} = Tr (\rho^{(p)}N_m)$ and 
$\langle N^2_m \rangle^{(p)} = Tr (\rho^{(p)} N^2_m)$, respectively, 
allow to obtain 
the  thermal intensity correlation function as follows:
\bea\label{thermcor}
(g^2)^{(p)}  =  \frac{\langle N^2_m \rangle^{(p)} - \langle   N_m \rangle^{(p)}}{\left(\langle   N_m \rangle^{(p)}\right)^{2}}.  
\ena
Then, the   thermal analogue of the Mandel parameter, given by
\bea\label{thermcor00}
 Q^{(p)}  =  \langle  N_m \rangle^{(p)}\left[(g^2)^{(p)}-1\right] 
\ena
is deduced.\\\\
In the illustrated examples, given an appropriate function $K_n^p(m), $   this formalism will be applied to
determine the concrete expressions
for the relations (\ref{thermal00})-(\ref{thermcor00}).
\section{GPAH-CS for classical orthogonal polynomials}
\ni In this section, the  Hermite, Laguerre, Jacobi polynomials and  hypergeometric functions are considered for the construction of GPAH-CS. 
\subsection{Coherent states for  associated Hermite and Laguerre  polynomials}
\ni The polynomials $\sigma$ and $\tau$ for Hermite and Laguerre polynomials  are: 
\bea \label{eq50}
\ba{llll}
 \sigma(x) = 1, & \quad \tau(x) = -x,& \quad \textrm{Hermite case}\\
  \sigma(x) = x, & \quad \tau(x) = \alpha +1 -x ,& \quad \textrm{Laguerre case.}
 \ea
 \ena
 In  both  cases $\sigma''=0$ and $\tau' = -1$. Then, $r_{m+n-k+1} = -(m+n-k)\sigma'' -\tau' = 1$.
  We have $f(r_{m+n-k+1})  = f(1) = \textrm{constant} = c$  for any analytical function  $f$.
 The eigenvalues, for any integer $l$, are  $\lambda_l = -\frac{1}{2}l(l-1)\sigma'' -l \tau' = l$, 
   so that $e_n = \lambda_{m+n}- \lambda_m  = n$. Then
 \bea \label{eq51}
 \varepsilon_n = e_1 e_2 \cdots \e_n = 1\cdot 2 \cdot 3 \cdots n = n!
 \ena
We obtain 
 \bea \label{eq52}
 \prod_{k=p}^{n + p -1}f(r_{m+n-k+1}) = c^n.
 \ena
The expansion coefficient (\ref{eq31}) follows from Eqs (\ref{eq51}) and (\ref{eq52}) as 
 \bea \label{eq53}
K_n^p(m) = {\Gamma(n+1)\over\sqrt{\Gamma(n+p+1)}}\, {1\over c^n}.
 \ena
 The normalization factor (\ref{eq32}) gives  here 
 \bea \label{eq54}
{\cal N}_p(|z|^2; m) = [\Gamma(p+1)\ _1F_1(1+p \ ; \ 1 \ ; \ |c z|^2)]^{-1/2},
 \ena
where  $_1F_1$ is the generalized  hypergeometric function. In terms 
of Meijer's G-function, the normalization factor is given by:
\bea \label{eq55}
{\cal N}_p(|z|^2; m) = \left[G_{1,2}^{1,1}\left(- {|c z|^2 } \lv
 \ba{ccc}-p & ; & \\ 0 & ; & 0 \ea  \right.  \right)\right]^{-1/2}.
\ena
  The explicit form of the GPAH-CS relative to Hermite and Laguerre polynomials, defined for 
any finite $|z|$,  follows from (\ref{eq30}):
  \bea \label{eq56}
\fl \lv z; m\rs_p = {1\over \sqrt{\dis \Gamma(p+1)\ _1F_1\left(p+1\ ; \ 1 \ ; \ |c z|^2\right)}} \, 
\sum_{n = 0}^\infty \sqrt{\dis \Gamma(n + p + 1)} \ {c^n z^n \over n!} \lv{n + p}\rs\pt
\ena 
For $p = 0$, we have ${\cal N}_0(|z|^2; m) = \exp{[-\half  |c z|^2]}$.
Performing the variable change  $ c z \longrightarrow z$, we 
recover, as expected,  for $p=0,$  the usual bosonic CS  \cite{Pere},
 \bea \label{eq57}
  |z; m \rangle  =  \exp\left[{-|z|^2\over2}\right]   \sum_{n=0}^\infty {z^n \over \sqrt{n!}}|n\rangle.
 \ena
Hence, the states  (\ref{eq56}), for $c =  1$, can be considered as the GPAH-CS for the harmonic oscillator. \\
\bei
\item[(i)]{\it Non-orthogonality}\\\\
The inner product  ${\cal P} = \,  _{p'} \li z'; m\right.\lv z; m\rs_p$ of two different GPACS-AHF
 $\lv z; m\rs_p $ and $\lv z'; m\rs_{p'}$ follows from (\ref{eq33}) as 
{\footnotesize 
\bea \label{eq58}
\fl  {\cal P}  & = & {\cal N}_{p'}(|z'|^2; m)\ {\cal N}_p(|z|^2; m){{(cz')}^\star}^{(p - p')}\ 
{\Gamma(p+1) \over \Gamma(p-p'+1)}\, _1F_1\left(p+1\ ; \ p - p' +1 \ ; \ |c|^2 z'^\star z\right)\\\label{eq59}
\fl & = & {\cal N}_{p'}(|z'|^2; m)\ {\cal N}_p(|z|^2; m){{(cz')}^\star}^{(p - p')}\ 
   G_{1,2}^{1,1}\left(\left.- {|c|^2 z'^\star z } \rv \ba{ccc}-p & ; & \\ 0 & ; & p' - p \ea    \right)
 \ena}
in terms of generalized hypergeometric and Meijer's G functions, respectively.\\
\begin{figure}[htbp]
\begin{center}
\includegraphics[width=11cm]{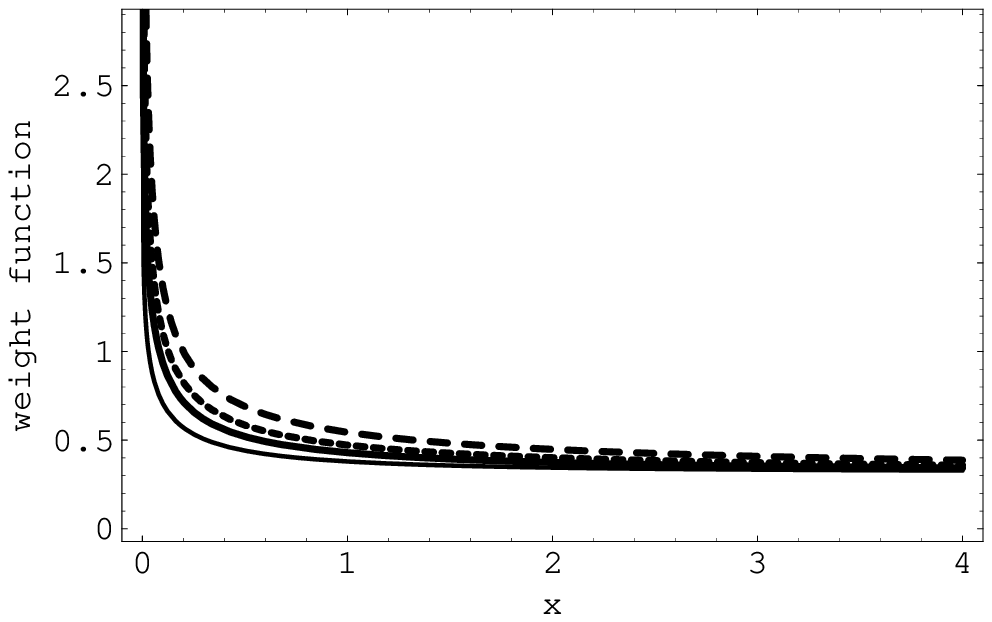}
\end{center}
\ni \caption[]{
Plots of the  weight function (\ref{eq64}) of  the
 GPAH-CS  (\ref{eq56}) versus  $ x = |z|^2$ with $c = 1$ and for different  values of 
 the photon added number $p$  with 
 $p = 1$ (thin solid  line), $p = 2$ (solid line), $p = 3$ (dot line), and $p = 4$ (dashed line). }
\end{figure}
\item[(ii)]{\it Overcompleteness}\\\\
\ni  The relation (\ref{eq39}) gives
\bea \label{eq60}
\int_0^\infty dx\ x^{n + p}\ g_p(x; m) & = & |K_n^p(m)|^2 = 
{\Gamma(n+1)^2\over \Gamma(n+p+1)} {1\over |c|^{2n}}
\ena
where the function $g_p$ is related to the measure $\omega_p$ as 
\bea \label{eq61}
\omega_p(x; m) = {x^p \over \pi}\, {1\over {\cal N}_p^2(x; m)}\, g_p(x; m).
\ena
Now,  performing the variable change $n + m \to s - 1$  transforms the   Eq. (\ref{eq60}) into: 
\bea \label{eq62}
\int_0^\infty dx \ x^{s-1}\ h_p(x; m) & = & \dis {1\over |c|^{2s}}\, {\Gamma(s-p)^2 \over \Gamma(s)} 
\ena
where $h_p(x; m) = g_p(x;m) |c|^{-2(p+1)}$. \ 
From  the Mellin-inversion theorem (\ref{eq41}), we deduce 
\bea \label{eq63}
h_p(x; m) & = & G_{1,2}^{2,0}\left(|c|^2 x\lv  \ba{rcl} & ;& 0 \\-p, -p & ; & \ea\right.\right).
\ena
The weight function  follows from (\ref{eq55}) and  (\ref{eq61}) as:
{
\bea \label{eq64}
\fl \omega_p(|z|^2 ; m) = {|c|^{2} \over \pi}\,
G_{1,2}^{1,1}\left(- {|c z|^2 } \lv \ba{ccc}-p & ; & \\ 0 & ; & 0 \ea  \right.  \right)
 G_{1,2}^{2,0}\left(\left.|cz|^2\rv  \ba{rcl} & ;& p \\0, 0 & ; & \ea\right)
\ena
where we use the multiplication formula of the Meijer's G-function \cite{Mathai}
\bea \label{eq64b}
x^\alpha  G_{m,n}^{p,q}\left(x\lv\ba{c} (a_p) \\ (b_q)\ea \right. \right) = G_{m,n}^{p,q}\left(x\lv\ba{c} (a_p + \alpha) \\ (b_q + \alpha)\ea \right. \right).
\ena
 The weight function (\ref{eq64}) is positive as confirmed by the Figure 1, where  it is represented for $p = 1, 2, 3, 4$. The curves show that the measure  has a singularity at $x = |z|^2 = 0$ and tends to zero for $x \to \infty$.}
For $p = 0$, we recover, as expected,  the   measure  $\omega_0(|z|^2; m) = \dis{|c|^2 \over \pi}$ obtained in our previous work \cite{Sodoga} for the corresponding ordinary coherent states.\\
\item[(iii)]{\it Photon number statistics}
\newline
\begin{figure}[htbp]
\begin{center}
\begin{minipage}{.45\textwidth}
\includegraphics[width=7cm]{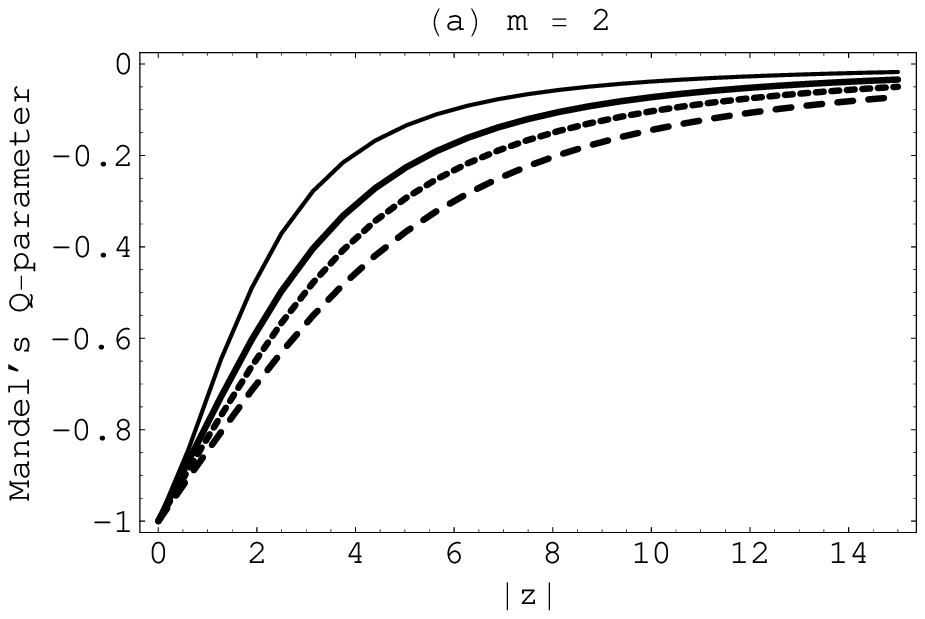}
\end{minipage} 
 \begin{minipage}{.45\textwidth}
\includegraphics[width=7cm]{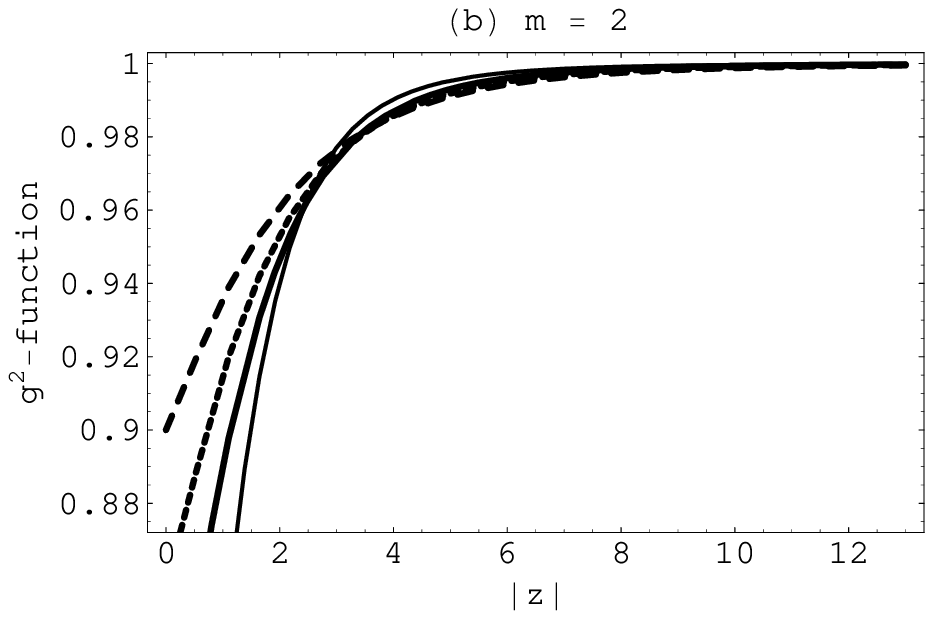}
\end{minipage}  
\end{center} 
\ni \caption[]{
 Plots of:  $(a)$ the Mandel Q-parameter  (\ref{eq72}) and (b) the second-order correlation function  (\ref{eq73}) of the GPAH-CS  (\ref{eq56}) versus   $|z|$ with the derivative order parameter 
$m = 2$ and  for various values of the photon-added number $p$   with  $p= 1$ (thin solid  line),  $p = 3$ (solid line), $p = 5$ (dot line) and $p = 8$ (dashed line).
} \label{Fig1_Hermite}
\end{figure}
\newline
\ni Taking into account   the expressions 
 (\ref{eq53}) and (\ref{eq54}) of the factors $K_n^p(m)$ and ${\cal N}_p(|z|^2; m)$,   
we obtain  the expectation values of Eqs. (\ref{eq46}) and (\ref{eq47}):
\bea \label{eq70}
\fl \li N_m \rs &=  & (m + p)\  {_2F_2(1+p,m+p+1;1,m+p ; |cz|^2) \over  _1F_1(1+p ; 1; |cz|^2)} \\\label{eq71}
\fl \li N_m^2 \rs &= &  (m + p)^2\  {_3F_3(1+p,m+p+1,m+p+1;1,m+p,m+p ; |cz|^2) \over _1F_1(1+p ; 1; |cz|^2)} \pt \ena 
\newline
\begin{figure}[htbp]
\begin{center}
\begin{minipage}{.45\textwidth}
\includegraphics[width=7cm]{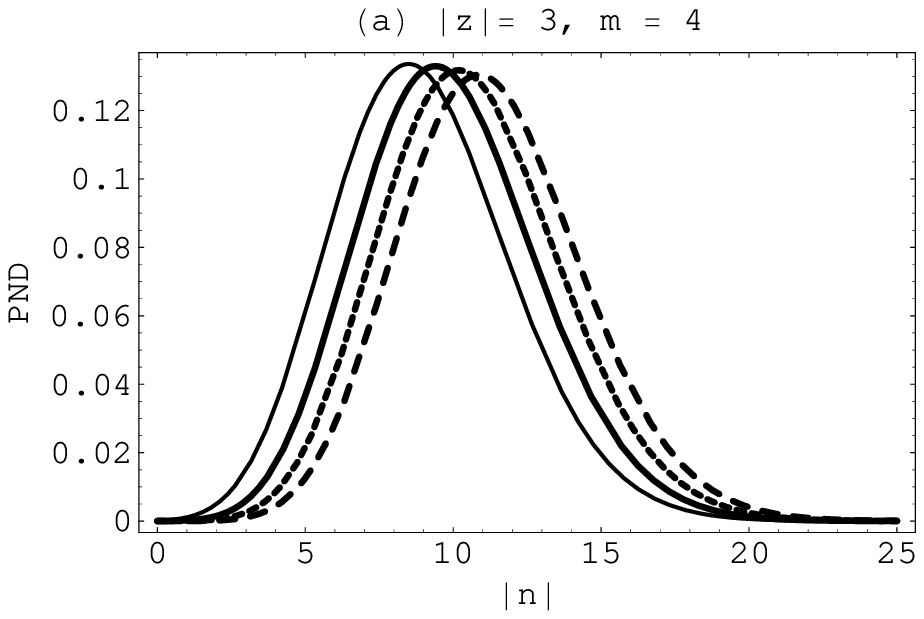}
\end{minipage} 
 \begin{minipage}{.45\textwidth}
\includegraphics[width=7cm]{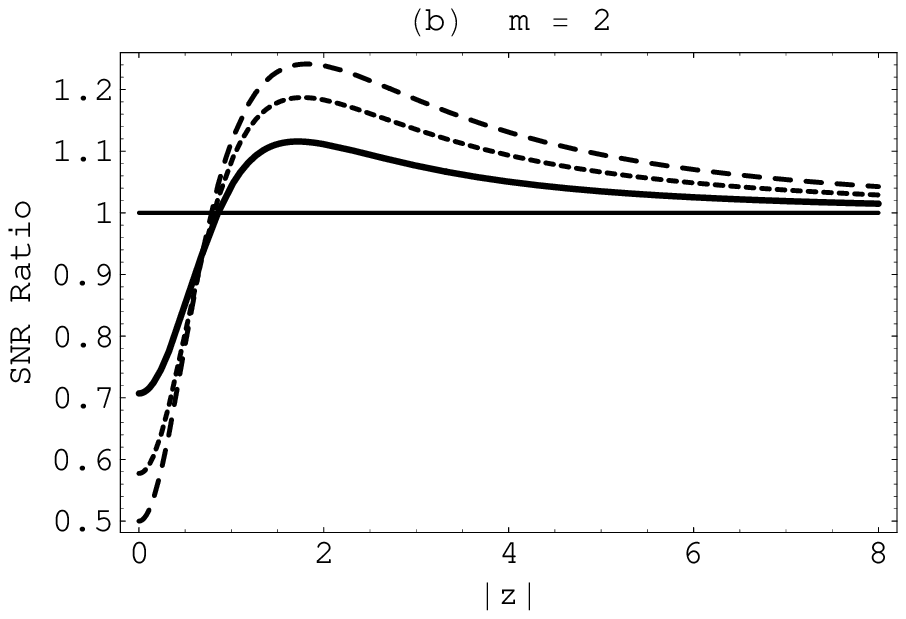}
\end{minipage}  
\end{center}
\ni \caption[]{
(a) Plot of the PND (\ref{PND_Hermite}) of the GPAH-CS  (\ref{eq56}) versus the photon number $n$, for $|z|=3, m = 4$ and different values of the photon-added number $p$ with  $p= 0$ (thin solid  line), $p = 1$ (solid line), $p = 2$ (dot line) and $p = 3$ (dashed line). \quad \\
 (b) 
Plot of the SNR ratios of GPAH-CS  (\ref{eq56}) 
 to the corresponding GAH-CS
 versus
  $|z|,$ for 
various values  of the photon-added number $p$ with  $p= 0$ (thin solid  line), $p = 1$ (solid line), $p = 2$ (dot line) $p = 3$ (dashed line).} \label{Fig2_Hermite}
\end{figure}
\ni Then, the Mandel Q-parameter is deduced as:
\bea \label{eq72}
Q = (m + p) \left[ {_3{\cal F}_3(|z|^2; m,p)\over_2{\cal F}_2 (|z|^2; m, p} -
     {_2{\cal F}_2(|z|^2; m, p)\over _1{\cal F}_1 (|z|^2; m, p)} \right] - 1.
\ena
The second order correlation function (\ref{eq44}) follows as 
\bea  \label{eq73}
g^2 = {_1{\cal F}_1(|z|^2; m, p)}\,{(m+p) _3{\cal F}_3(|z|^2; m, p) - _2{\cal F}_2(|z|^2; m, p)
 \over (m+p) _2{\cal F}_2(|z|^2; m, p)^2}
\ena
where $_1{\cal F}_1$, $_2{\cal F}_2$ and  $_3{\cal F}_3$ are the generalized hypergeometric functions:
\beano
\fl _1{\cal F}_1 (|z|^2; m, p) & = & _1F_1(1+p ; 1; |cz|^2)\\
\fl _2{\cal F}_2(|z|^2; m, p) & = &  _2F_2(1+p,m+p+1;1,m+p ; |cz|^2)\\
\fl _3{\cal F}_3(|z|^2; m, p) & = & _3F_3(1+p,m+p+1,m+p+1;1,m+p,m+p ; |cz|^2) \pt
\enano
The PND (\ref{PND}) reads as 
\bea\label{PND_Hermite}
 {\cal P}_n^p(|z|^2; m) = {\Gamma(n+1) \over \Gamma(p+1)\, _1F_1(1+p ; 1; |cz|^2)}\, 
{|cz|^{2(n-p)} \over (n-p)!^2}
\ena
which gives for $p = 0$
\bea\label{PND_Hermitep=0}
 {\cal P}_n^p(|z|^2; m) = e^{-|cz|^{2}}
{|cz|^{2n} \over n!}.
\ena
\ni Finally the SNR  (\ref{SNR}) gives 
{\footnotesize
\bea \label{SNR_Hermite}
\fl \sigma^p(|z|^2; m) = (m + p){_2{\cal F}_2(|z|^2; m, p) - p _1F_1(1+p ; 1; |cz|^2) \over 
(m+ p) \sqrt{_3{\cal F}_3(|z|^2; m,p) _1{\cal F}_1(|z|^2; m,p) - (_2{\cal F}_2(|z|^2; m,p))^2}}.\\
\ena}
\item[(iv)] {\it  Thermal statistics}\\\\
The relative normalized density operator and the partition function are  provided as 
\bea\label{dmatrix00}
 \fl\rho^{(p)}  &=& \frac{1}{Z}\sum_{n = 0} e^{-\beta n} |n+p\rangle \langle n+p|\cr
\fl &=&\frac{1}{\bar{n}_o+1} \sum_{n = 0} \left(\frac{\bar{n}_o}{\bar{n}_o+1}\right)^n |n+p\rangle \langle n+p|, 
\qquad  Z =  \frac{1}{1-e^{-\beta}} :=  \bar{n}_o+1  
\ena
respectively, with $\bar{n}_o =(e^\beta -1)^{-1}$ corresponding to the thermal expectation value of the number operator, i.e. 
the Bose-Einstein distribution  function for oscillators with angular frequency $\omega =1, $ where $\hbar =1.$  \\
 From (\ref{eq56}) and (\ref{dmatrix00}), we get, in terms of generalized hypergeometric functions,  the $Q$-distribution
\bea\label{dmatrix01}
_{p}\langle z;m| \rho^{(p)} |z;m\rangle= (\bar{n}_o+1)
\frac{_1F_1\left(p+1\ ; \ 1 \ ; \  |c z|^2 e^{-\beta} \right)}{_1F_1\left(p+1\ ; \ 1 \ ; \ |c z|^2\right)}
\ena
and in terms of Meijer's G functions,
{\footnotesize
\bea\label{dmatrix02}
_{p}\langle z;m| \rho^{(p)} |z;m\rangle_p = (\bar{n}_o+1)\frac{G_{1,2}^{1,1}\left(- { |cz|^2 e^{-\beta}} \lv
 \ba{ccc}-p & ; & \\ 0 & ; & 0 \ea  \right.  \right)}{G_{1,2}^{1,1}\left(- {|c z|^2 } \lv
 \ba{ccc}-p & ; & \\ 0 & ; & 0 \ea  \right.  \right)}.
\ena}
After performing the angular integration and taking $x =|c z|^2, $ the condition (\ref{thermal02}) leads to 
{\footnotesize
\bea\label{dmatrix03}
\mbox{Tr}  \rho^{(p)}  = \frac{1}{Z}\int_{0}^{\infty} dx \, x^p \, 
 G_{1,2}^{1,1}\left({x } \lv
 \ba{ccc}-p & ; & \\ 0 & ; & 0 \ea  \right.  \right)
G_{1,2}^{1,1}\left(- {x e^{-\beta}} \lv
 \ba{ccc}-p & ; & \\ 0 & ; & 0 \ea  \right.  \right).
\ena}
Thus, using   the properties of the integral of Meijer's G-functions products provides the partition function expression
\bea
 Z =  \frac{1}{1-e^{-\beta}} =  \bar{n}_o+1.  
\ena
From (\ref{thermal03}),   using the result 
\bea\label{dmatrix04}
\langle n+p| \rho^{(p)} |n+p\rangle  =  \frac{1}{\bar{n}_o+1}\left(\frac{\bar{n}_o}{\bar{n}_o+1}\right)^n
\ena
we obtain the following integration  equality
{\footnotesize
\bea\label{dmatrix05}
\fl\frac{1}{\bar{n}_o+1}\left(\frac{\bar{n}_o}{\bar{n}_o+1}\right)^n \frac{1}{|c|^{2(n+p+1)}}\frac{\Gamma(n+1)^2}{\Gamma(n+p+1)} 
= \int_{0}^{\infty} dx \, x^{n+p} P(x) G_{1,2}^{2,0}\left(\left.|c|^2 x\rv  \ba{rcl} & ;& 0 \\-p, -p & ; & \ea\right).\nonumber
\ena}
After performing  the exponent change  $ n+p = s-1 $  of $x = |z|^2, $ in order to get to the Stieltjes moment problem, 
we arrive at the $P$-function  obtained as 
{\footnotesize
\bea\label{thermal05}
P(|z|^2) =  \frac{1}{\bar{n}_o}\left(\frac{\bar{n}_o+1}{\bar{n}_o}\right)^p 
\frac{G_{1,2}^{2,0}\left(\left.\frac{\bar{n}_o+1}{\bar{n}_o} |c z|^2\rv  \ba{rcl} & ;& 0 \\-p, -p & ; & \ea\right)}
{G_{1,2}^{2,0}\left(\left. |c z|^2\rv  \ba{rcl} & ;& 0 \\-p, -p & ; & \ea\right)}.
\ena} 
The resulting $P$-function (\ref{thermal05}) obeys the normalized to unity condition (\ref{thermal04}).
Thereby, the diagonal representation of the normalized density operator in terms of the  GPAH-CS
projector (\ref{thermal03}),  in this case, is 
{\footnotesize
\bea
\fl \rho^{(p)}  
 =  \frac{1}{\bar{n}_o}\left(\frac{\bar{n}_o+1}{\bar{n}_o}\right)^p  \int_{\IC} d^2 z \, \omega_p(|z|^2; m) |z;m\rangle_p
\frac{G_{1,2}^{2,0}\left(\left.\frac{\bar{n}_o+1}{\bar{n}_o} |c z|^2\rv  \ba{rcl} & ;& 0 \\-p, -p & ; & \ea\right)}
{G_{1,2}^{2,0}\left(\left.|c z|^2\rv  \ba{rcl} & ;& 0 \\-p, -p & ; & \ea\right)} \, _{p}\langle z;m|.
\ena               
}
With the relations (\ref{eq70}), (\ref{eq71}) and the definition (\ref{obaverage00}),   the 
pseudo-thermal expectation values of the number operator and of its 
  square  are obtained as 
\bea\label{numbtherm00} 
\fl\langle  N_m \rangle^{(p)} &=& Tr ( \rho^{(p)}   N_m) =   \frac{1}{\bar{n}_o+1}(m+p)\sum_{n=0}^{\infty}\left(1 + \frac{n}{m+p+1}\right)
\left(\frac{\bar{n}_o}{\bar{n}_o+1}\right)^n \cr
\fl\langle   N^2_m \rangle^{(p)} &=& Tr ( \rho^{(p)}    N^2_m) = 
\frac{1}{\bar{n}_o+1}(m+p)^2\sum_{n=0}^{\infty}\left(1 + \frac{n}{m+p+1}\right)^2
\left(\frac{\bar{n}_o}{\bar{n}_o+1}\right)^n.
\ena
Using (\ref{numbtherm00}),   
the  thermal intensity correlation function (\ref{thermcor}), is obtained as 
\bea\label{numbtherm01}
(g^2)^{(p)}  =  1 - \frac{\left[(m+p+1)^2 +  \bar{n}_o\right]- \bar{n}_o^2(m+p)}{(m+p)\left[(m+p+1)+  \bar{n}_o \right]^2}.  
\ena
\ni
Thereby, the  thermal analogue of the Mandel parameter (\ref{thermcor00})
is
\bea\label{numbtherm03}
Q^{(p)}  = \frac{\bar{n}_o^2(m+p)-\left[(m+p+1)^2 +  \bar{n}_o\right]}{(m+p+1)\left[(m+p+1)+  \bar{n}_o \right]}.
\ena
\eni
\subsection{Coherent states for associated Jacobi polynomials and hypergeometric functions}
\ni The polynomial coefficients $\sigma$ and $\tau)$ for Jacobi  and hypergeometric cases are: 
\bea \label{eq75}
\fl \ba{llll}
 \sigma(x) = 1- x^2, & \quad \tau(x) = (\beta - \alpha) -(\alpha + \beta + 2)x,& \quad \textrm{ for Jacobi polynomials}\\
  \sigma(x) = x(1- x), & \quad \tau(x) = (\alpha + 1) -(\alpha + \beta + 2)x ,& \quad \textrm{for Hypergeometric functions}\\
 \ea
 \ena
 In  both the  cases,  $\sigma''=-2$ and $\tau' = -(\alpha + \beta + 2) = -\mu$.
 The eigenvalues, for any integer $l$, are 
\bea \label{eq76}
\fl  \lambda_l = -\frac{1}{2}l(l-1)\sigma'' -l \tau' = l(l + \mu -1), \quad
  e_n = \lambda_{m+n}- \lambda_m  = n(n+2m+\mu -1).
\ena
 Then the quantities  $\varepsilon_n$, for any integer  $n$, can be read as: 
 \bea \label{eq77}
  \varepsilon_n = e_1 e_2 \cdots \e_n = \Gamma(n+1)\,{\Gamma(n+ 2 m + \mu)\over \Gamma(2 m + \mu)}. 
 \ena
The variable $r_{m+n-k+1}$ of the functional $f$ in the expansion coefficient (\ref{eq31}) is written as:
\bea \label{eq78}
r_{m+n-k+1} = -(m+n-k)\sigma'' -\tau' = 2(m+n-k) + \mu 
\ena 
The explicit form of the GPAH-CS depends on the expansion coefficient $K_n^p(m)$  which is  defined in terms of the functional $f$. 
In the following, we adopt  some functionals close to those used in \cite{Aleixo}. 
\subsubsection{Case where $f$ is constant} \quad \\\\
First, we consider that   the analytical  function $f$ is  a constant $c$. 
Then, we have $f(r_{m+n-k+1}) = \textrm{constant} = c$, and
\bea \label{eq79}
 \prod_{k=p}^{n + p -1}f(r_{m+n-k+1}) = c^n.
 \ena
The expansion coefficient (\ref{eq31})  from Eqs (\ref{eq77}) and (\ref{eq79}) becomes
 \bea \label{eq80}
K_n^p(m) = {\Gamma(n+1)\Gamma(n+ 2 m + \mu) 
\over \sqrt{\Gamma(2 m + \mu)  \Gamma(n + p +1)  \Gamma(n + p + 2 m + \mu)}}\, {1\over c^n}.
 \ena
Without loss of generality in the sequel, we take $c = 1$. Then, the expansion coefficient gives
 \bea \label{eq81}
K_n^p(m) = {\Gamma(n+1)\Gamma(n+ 2 m + \mu) 
\over \sqrt{\Gamma(2 m + \mu)  \Gamma(n + p +1)  \Gamma(n + p + 2 m + \mu)}}.
 \ena
For $p = 0$, this reduces to
 \bea \label{eq82}
K_n^0(m) = \sqrt{\dis {\Gamma(n+1)\Gamma(n+ 2 m + \mu) 
\over \Gamma(2 m + \mu) }},
 \ena
and we recover the result of the corresponding CS, obtained in \cite{Sodoga}.\\\\
 The normalization factor (\ref{eq32}) is formulated in terms of the generalized hypergeometric function $_2F_3$
{\footnotesize
\bea \label{eq83}
\fl {\cal N}_p(|z|^2; m) = \left[{\Gamma(p+1)  \Gamma(2 m + p + \mu) \over  \Gamma(2 m + \mu)} 
 \ _2F_3(1+p, 2m + p + \mu \ ; \ 1, 2 m + \mu, 2 m + \mu \ ; \ |z|^2)\right]^{-1/2},
 \ena}
or in terms of Meijer G-function
{\small
\bea \label{eq84}
\fl {\cal N}_p(|z|^2; m) = \left[\Gamma(2 m + \mu)\, G_{2,4}^{1,2}\left(- {|z|^2 } \lv
 \ba{l}-p, 1 -2 m -p - \mu  \\ 0 , 0, 1 -2 m - \mu, 1 -2 m - \mu  \ea   \right. \right)\right]^{-1/2}.
\ena}
where from now and in the sequel we use   the more compact  notation of long Meijer's G-function.
\bea \label{eq84b}
 G_{p,q}^{m,n} 
\left( x\lv \ba{c} 
a_1,... , a_p \\
b_1,... , b_q 
\ea \right.\right)\pt
\ena
For $p = 0$, it is  reduced to the result  obtained in \cite{Sodoga} for
 the  corresponding  CS, expressed with
 \bea \label{eq85}
\fl {\cal N}_0(|z|^2; m) = [_0F_1(2m + \mu, |z|^2)]^{-1/2} = [|z|^{1-\mu - 2m}I_{2 m + \mu - 1}(2|z|) \Gamma(2 m + \mu)]^{-1/2}.
 \ena 
 $_0F_1$ is the confluent hypergeometric function and  $I_\nu$- the modified Bessel function of the first kind.\\\\
The explicit form of the GPAH-CS relative to Jacobi and Hypergeometric  polynomials, defined for any finite $|z|$,  follows from (\ref{eq30}):
{
\bea \label{eq86}
\fl \lv z; m\rs_p = {\cal N}_p(|z|^2; m) \, 
\sum_{n = 0}^\infty \sqrt{\dis { \Gamma(2 m + \mu)  \Gamma(n + p +1)  \Gamma(n + p + 2 m + \mu)
\over  \Gamma(n+ 2 m + \mu)^2} }\ {z^n \over n!} \lv n + p\rs \pt
\ena}
For $p = 0$, we  recover the corresponding CS  obtained in \cite{Sodoga}, 
  \bea \label{eq87},
 \fl  |z;m\rangle = \frac{1}{\sqrt{_0F_1(2m+ \mu, |z|^2)}}\sum_{n=0}^\infty \frac{z^n}
  {\sqrt{\Gamma(n+1)(2m + \mu)_n}}|n\rangle , \qquad |z| < \infty
  \ena
as a particular case.
\bei
\item[(i)]{\it Non-orthogonality}\\ \\
The inner product  ${\cal P} = \,  _{p'} \li z'; m\right.\lv z; m\rs_p$ of two different GPAH-CS
 $\lv z; m\rs_p $ and $\lv z'; m\rs_{p'}$, for Jacobi polynomials  and hypergeometric functions, follows from (\ref{eq33}) as 
{
\bea \label{eq90}
 \fl  {\cal P} & = & {\cal N}_{p'}(|z'|^2; m)\ {\cal N}_p(|z|^2; m){{z'}^\star}^{(p - p')}\ 
{\Gamma(p+1) \Gamma(p + 2m +\mu) \over \Gamma(p-p'+1) \Gamma(p - p'+2m +\mu)}\times \cr
\fl  &  &  _2F_3\left(p+1, 2 m + p + \mu\ ; \ p - p' +1, 2m +\mu, 2m +\mu + p - p' \ ; \ z'^\star z\right)
 \ena}
yielding, for $p = p' = 0$, 
\bea \label{eq91}
 _{0} \li z'; m\right.\lv z; m\rs_0 = \frac{_0F_1(2m+ \mu, z'^\star z)}{\sqrt{_0F_1(2m+ \mu, |z'|^2)\, _0F_1(2m+ \mu, |z|^2)}}.\\ \nonumber
\ena
\begin{figure}[htbp]
\begin{center}
\includegraphics[width=11cm]{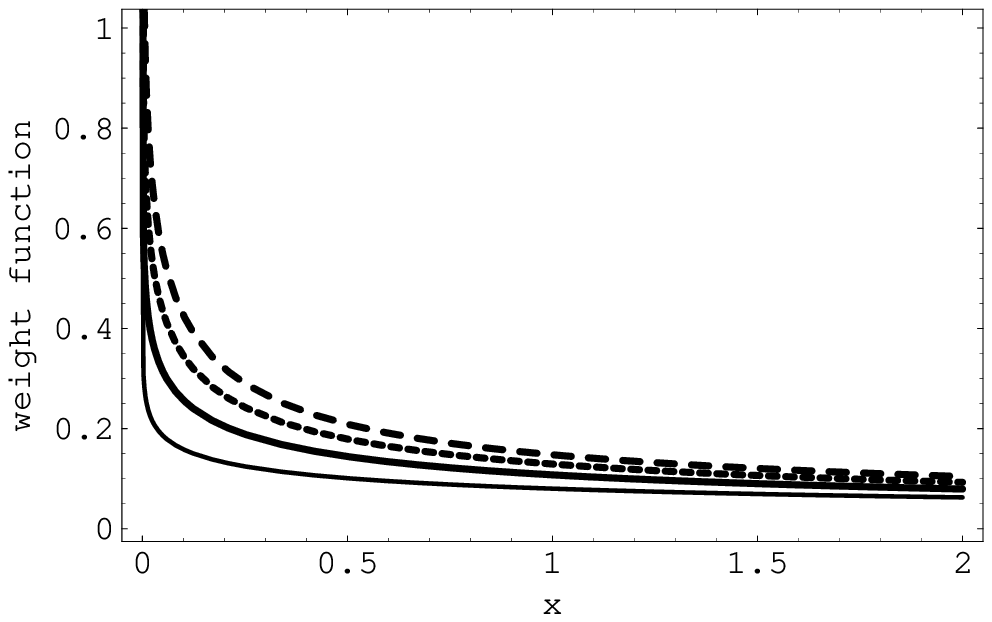}
\end{center}
\ni \caption[]{
Plots of the  weight function (\ref{eq95}) of  the
 GPAH-CS (\ref{eq86}) versus  $x = |z|^2$  with  parameters 
$m = 4, \mu =1.5$ and  for different  values of 
 the photon added number $p$  with 
 $p = 1$ (thin solid  line), $p = 2$ (solid line), $p = 3$ (dot line), and $p = 4$ (dashed line). } 
\end{figure}
\item[(ii)]{\it Overcompleteness}\\\\
\ni  The relation (\ref{eq39}) now engenders for the  Jacobi polynomials and  hypergeometric functions:
\bea \label{eq92}
\fl \int_0^\infty dx\ x^{n + p}\ g_p(x; m) & = &   
{\Gamma(n+1)^2 \Gamma(n+ 2 m + \mu)^2 \over \Gamma(2 m + \mu) \Gamma(n+p+1) \Gamma(n+ p +2 m + \mu)}
\ena
where the function $g_p$ is related to the measure $\omega_p$ by (\ref{eq61}).
Provided the variable change $n + m \to s - 1$, it becomes: 
\bea \label{eq93}
\int_0^\infty dx \ x^{s-1}\ h_p(x; m) & = & 
\dis  {\Gamma(s-p)^2  \Gamma(s-p - 1 + 2 m + \mu)^2\over \Gamma(s) \Gamma(s - 1 + 2 m + \mu)} 
\ena
where $h_p(x; m) = g_p(x;m)  \Gamma(2 m + \mu)$. \ 
From  the Mellin-inversion theorem (\ref{eq41}),  we deduce 
{\small
\bea \label{eq94}
\fl h_p(x; m) & = & G_{2,4}^{4,0}\left(x \lv
  \ba{l}  0, -1 + 2 m + \mu \\ -p, -p, -1 -p + 2 m + \mu, -1 -p + 2 m + \mu  \ea \right. \right).
\ena}
The weight function  follows from (\ref{eq61}) and (\ref{eq84}) as
{
{\footnotesize 
\bea \label{eq95}
\fl \omega_p(|z|^2 ; m) & = & {1 \over \pi}\, 
 G_{2,4}^{1,2}\left(- {|z|^2 } \lv
 \ba{l}-p, 1 -2 m -p - \mu  \\ 0 , 0, 1 -2 m - \mu, 1 -2 m - \mu  \ea   \right. \right)\times  \cr
\fl & & \times G_{2,4}^{4,0}\left(|z|^2\lv
  \ba{l}  p, -1 + p+  2 m + \mu \\ 0, 0, -1 + 2 m + \mu, -1  + 2 m + \mu  \ea \right.\right)
\ena}
using the multiplication formula of the Meijer's G-function (\ref{eq64b}). 
The weight function (\ref{eq95}) is positive for the parameter $\mu > 0$ as shown on the representations in Figure 4 with  $\mu =1.5$ and for  different values of the photon-added number $p = 1, 2, 3, 4$.
Figure 4 also shows  that  the weight function (\ref{eq95}) presents a singularity at $x = 0$  and tends to zero for $x \to \infty.$}
For $p = 0$, the Meijer G function $G_{2,4}^{4,0}$ reduces to $2|z|^{-1 +\mu + 2m}\,K_{2 m + \mu - 1}(2|z|),$ and taking in to account (\ref{eq85}), we  retrieve the   measure obtained in  \cite{Sodoga} for the corresponding ordinary coherent states,   
\beano
\omega_0(|z|^2; m) & = &{2\over \pi}I_{2 m + \mu - 1}(2|z|)\,K_{2 m + \mu - 1}(2|z|),
\enano
where $I_\nu$  and $K_\nu$   are the   modified Bessel functions of first and second kind, respectively.\\
\item[(iii)]{\it Photon number statistics}
\newline
\begin{figure}[htbp]
\begin{center}
\begin{minipage}{.45\textwidth}
\includegraphics[width=7cm]{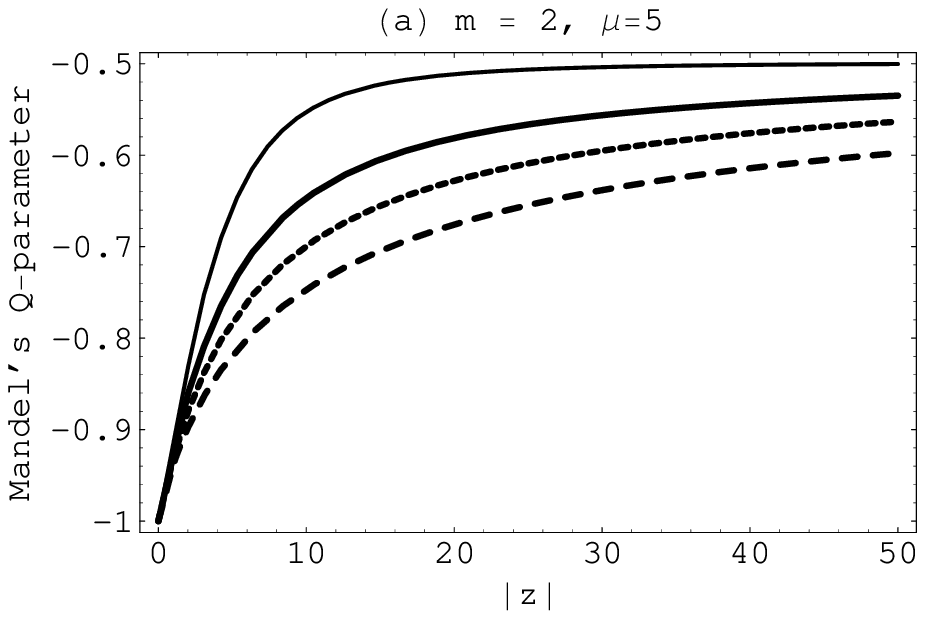}
\end{minipage} 
 \begin{minipage}{.45\textwidth}
\includegraphics[width=7cm]{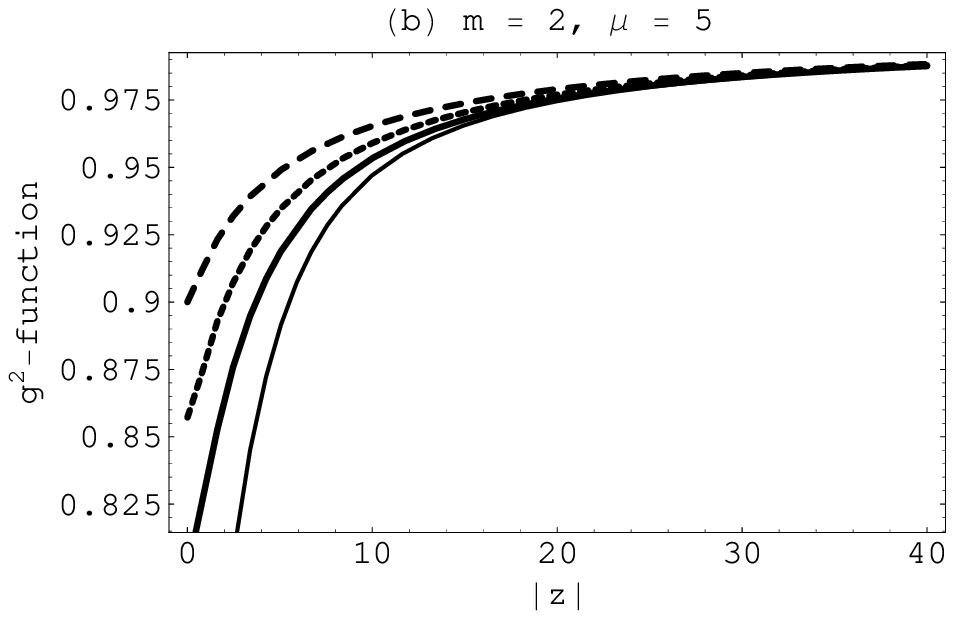}
\end{minipage}  
\end{center}
\ni \caption[]{
 Plots of: $(a)$ the Mandel Q-parameter  (\ref{eq98}) and (b) the second order correlation function (\ref{eq99}) of the GPAH-CS (\ref{eq86}) versus  $|z|$ with  parameters 
$m = 2, \mu =5$ and  for various values of the photon-added number $p$ with  $p= 1$ (thin solid  line),  $p = 3$ (solid line), $p = 5$ (dot line) and $p = 8$ (dashed line).
} \label{Fig1_Jacobiex1}
\end{figure}
\newline
\begin{figure}[htbp]
\begin{center}
\begin{minipage}{.45\textwidth}
\includegraphics[width=7cm]{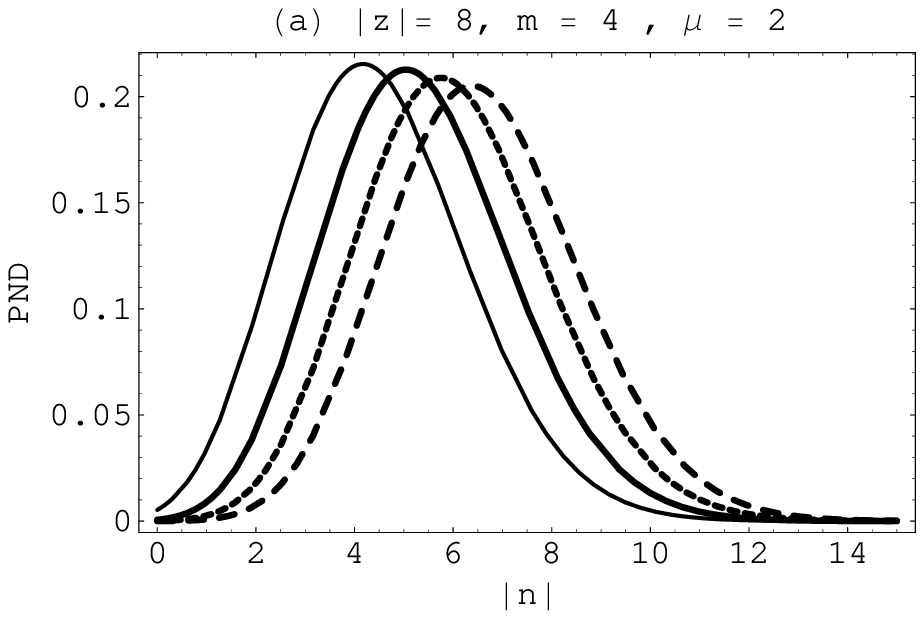}
\end{minipage} 
 \begin{minipage}{.45\textwidth}
\includegraphics[width=7cm]{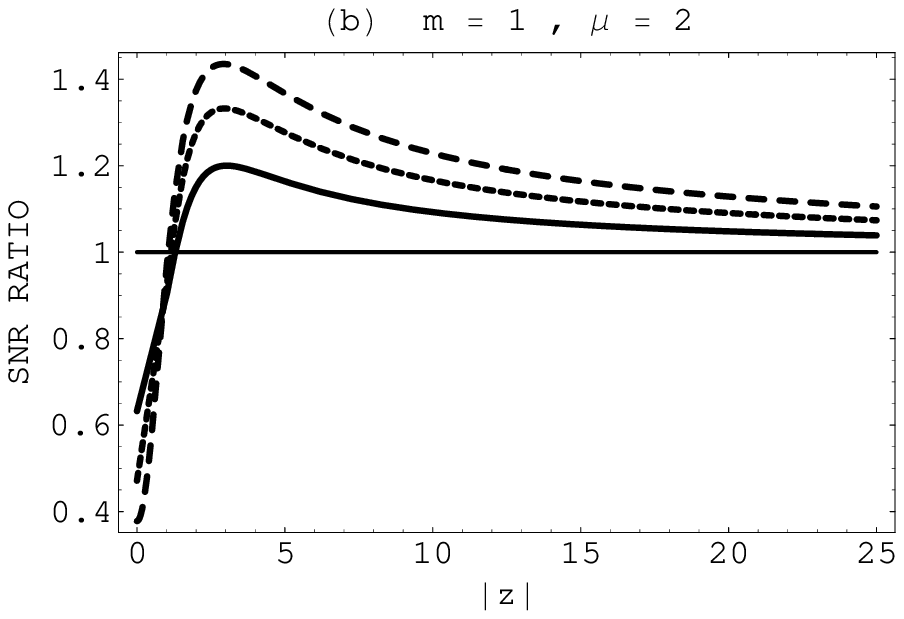}
\end{minipage}  
\end{center}
\ni \caption[] 
{
(a) Plot of the  PND (\ref{PND_Jacobi1}) of the GPAH-CS  (\ref{eq86}) versus the photon number $n$, with the parameters $m = 4 ,  \mu =2$ and for different values of the photon-added number $p$ with $p= 0$ (thin solid  line),
 $p = 1$ (solid line), $p = 2$ (dot line) and $p = 3$ (dashed line). \quad \\
 (b) 
Plot of the SNR (\ref{SNR_Jacobi1}) ratios of GPAH-CS (\ref{eq86})  and the corresponding GAH-CS versus  $|z|$, with the parameters $m = 1 ,  \mu =2,$ for 
various values  of the photon-added number $p$ with $p= 0$ (thin solid  line), $p = 1$ (solid line), $p = 2$ (dot line) and $p = 3$ (dashed line)} \label{Fig2_Jacobi1}
\end{figure}
\newline
Taking into account   the expressions 
 (\ref{eq81}) and (\ref{eq83}) of the factors $K_n^p(m)$ and ${\cal N}_p(|z|^2; m)$,   
we obtain  the expectation values (\ref{eq46},\ref{eq47}):
{\small
\bea \label{eq96}
\fl \li N_m \rs &=  & (m + p)\  {_3F_4\left(\ba{lcr}1+p,m+p+1, p+2m+ \mu &; & \\
1,m+p, 2 m + \mu, 2 m + \mu & ; & |z|^2\ea\right) \over
  _2F_3\left(\ba{lcr}1+p,2m +p +\mu &; & \\1,  2 m + \mu, 2 m + \mu &; & |z|^2\ea \right)} \\\label{eq97}
\fl \li N_m^2 \rs &= &  (m + p)^2\ 
 {_4F_5\left(\ba{lcl}1+p,m+p+1,m+p+1, p+2m+ \mu &;&\\
                                    1,m+p,m+p,  2 m + \mu, 2 m + \mu & ; & |z|^2 \ea \right) \over
 _2F_3\left(\ba{lcr}1+p,2m +p +\mu &; &\\ 1,  2 m + \mu, 2 m + \mu &; & |z|^2\ea \right)} \pt \ena }
where we adopt from now and in the sequel the following notation 
\bea\label{pfqlong}
_pF_q\left(\ba{lcr} a_1, \ldots, a_p  &;&\\ 
                    b_1, \ldots, b_q  &;& x\ea\right).
\ena
for long $_pF_q$ hypergeometric.
Then, the Mandel Q-parameter can be inferred as:
\bea \label{eq98}
Q = (m + p) \left[ {_4{\cal F}_5(|z|^2; m,p)\over _3{\cal F}_4 (|z|^2; m, p} -
     {_3{\cal F}_4(|z|^2; m, p)\over  _2{\cal F}_3 (|z|^2; m, p)} \right] - 1
\ena
while the second order correlation function (\ref{eq44}) yields 
\bea  \label{eq99}
g^2 = { _2{\cal F}_3(|z|^2; m, p)}\,{(m+p)_4{\cal F}_5(|z|^2; m, p) -_3{\cal F}_4(|z|^2; m, p)
 \over (m+p)_3{\cal F}_4(|z|^2; m, p)^2}
\ena
where $_2{\cal F}_3$, $_3{\cal F}_4$ and  $_4{\cal F}_5$ are the generalized hypergeometric functions:
\beano
\fl  _2{\cal F}_3 (|z|^2; m, p) & = & _2F_3\left(\ba{lcr}1+p, 2m +p +\mu &;&
                                 \\ 1,  2 m + \mu, 2 m + \mu &; & |z|^2\ea \right)\\
\fl _3{\cal F}_4 (|z|^2; m, p) & = & _3F_4\left(\ba{lcr}1+p,m+p+1, p+2m+ \mu &;& \\ 
                                  1,m+p, 2 m + \mu, 2 m + \mu &;& |z|^2\ea \right) \\
\fl _4{\cal F}_5(|z|^2; m, p) & = & _4F_5\left(\ba{lcl}1+p,m+p+1,m+p+1, p+2m+ \mu &;&\\
                                    1,m+p,m+p,  2 m + \mu, 2 m + \mu & ; & |z|^2 \ea \right)  \pt
\enano
The PND (\ref{PND}) reads as 
{\footnotesize
\bea\label{PND_Jacobi1}
 \fl {\cal P}_n^p(|z|^2; m) = {\Gamma(2m+\mu)^2\Gamma(n+2m+\mu) \Gamma(n+1) \over 
\Gamma(p+2m+\mu)\Gamma(p+1) \Gamma(n-p +2m+\mu)^2\, _2{\cal F}_3 (|z|^2; m, p) }\, 
{|z|^{2(n-p)} \over (n_p)!^2}.
\ena}
This gives, for $p = 0$, the result
\bea\label{PND_Jacobi1_p=0}
 \fl {\cal P}_n^p(|z|^2; m) = {\Gamma(2m+\mu) \over 
\Gamma(n+2m+\mu)\, _0F_1 (2m+ \mu, |z|^2) }\, 
{|z|^{2n} \over n!}
\ena
corresponding to the conventional GAH-CS PND.
\ni Finally the SNR  (\ref{SNR}) gives 
{\footnotesize
\bea \label{SNR_Jacobi1}
\fl \sigma^p(|z|^2; m) = {(m + p)\, _3{\cal F}_4(|z|^2; m, p) - p\  _2F_3(1+p ; 1; |z|^2) \over 
(m+ p) \sqrt{_4{\cal F}_5(|z|^2; m,p) _2{\cal F}_3(|z|^2; m,p) - (_3{\cal F}_4(|z|^2; m,p))^2}}.
\ena}
\item[(iv)] {\it Thermal statistics}\\\\
 Consider the normalized density operator expression 
 \bea\label{dmatrix06}
 \rho^{(p)}   =  \frac{1}{Z}\sum_{n = 0} e^{-\beta e_n} |n+p\rangle \langle n+p|\cr
\ena 
 in which the exponent $ \beta e_n$ is recast   as follows
 \bea
 \beta e_n =   \beta \left[n^2+(2m + \mu -1)n\right]   = A n^2 - B_{m,\mu}n
 \ena
 where $ A = \beta, \, B_{m,\mu} = -\beta  (2m + \mu -1). $ Then, the energy exponential can  be expanded in the 
 power series,   (see for e.g., \cite{popov01}) such that
{
\bea
 \fl e^{-\beta e_n}=  e^{-A n}\left[\sum_{k= 0}^{\infty} \frac{(B_{m,\mu})^k}{k!} n^{2k}\right] &= & 
 \left\{\sum_{k= 0}^{\infty} \frac{(B_{m,\mu})^k}{k!} \left(\frac{d}{dA}\right)^{2k}\right\} \left(e^{-A}\right)^n 
\cr &= & 
\mbox{exp}\left[B_{m,\mu}\left(\frac{d}{dA}\right)^{2} \right]\left(e^{-A}\right)^n.
 \ena}
 Thereby, 
\bea\label{dmatrix07}
 \rho^{(p)}  &=& \frac{\mbox{exp}\left[B_{m,\mu}\left(\frac{d}{dA}\right)^{2} \right]}{Z}\sum_{n = 0}^{\infty} 
\left(e^{-A}\right)^n |n+p\rangle \langle n+p|.  
\ena
 From (\ref{eq86}) and (\ref{dmatrix07}), we get, in terms of generalized hypergeometric functions,  the $Q$-distribution or Husimi 
distribution
{\footnotesize
\bea 
\fl _{p}\langle z;m| \rho^{(p)} |z;m\rangle_p = \frac{\mbox{exp}\left[B_{m,\mu}\left(\frac{d}{dA}\right)^{2} \right]}{Z}
\frac{_2F_3\left(\ba{lcr}1+p, 2m + p + \mu \ &;& \\
 \ 1, 2 m + \mu, 2 m + \mu \ &;& |z|^2 e^{-A}\ea\right)}
{_2F_3\left(\ba{lcr}1+p, 2m + p + \mu \ &; & \\ 1, 2 m + \mu, 2 m + \mu \ &; & \ |z|^2\ea \right)}\nonumber
\ena}
and in terms of Meijer's G functions,
{\footnotesize
\bea 
 \fl && _{p}\langle z;m| \rho^{(p)} |z;m\rangle_p = \frac{\mbox{exp}\left[B_{m,\mu}\left(\frac{d}{dA}\right)^{2} \right]}{Z}
\frac{ G_{2,4}^{1,2}\left(- {|z|^2 e^{-A}} \lv
 \ba{l}-p, 1 -2 m -p - \mu  \\ 0 , 0, 1 -2 m - \mu, 1 -2 m - \mu  \ea   \right. \right)}
 { G_{2,4}^{1,2}\left(- {|z|^2 } \lv
 \ba{l}-p, 1 -2 m -p - \mu  \\ 0 , 0, 1 -2 m - \mu, 1 -2 m - \mu  \ea   \right. \right)}.
\ena}
The   angular integration achieved, taking    $x = |z|^2, $ the condition (\ref{thermal02}) supplies 
{\footnotesize
\bea 
 \fl\mbox{Tr}  \rho^{(p)} & = & \frac{\mbox{exp}\left[B_{m,\mu}\left(\frac{d}{dA}\right)^{2} \right]}{Z} \int_{0}^{\infty} dx \, x^p 
G_{2,4}^{1,2}\left(- {x e^{-A}} \lv
 \ba{l}-p, 1 -2 m -p - \mu  \\ 0 , 0, 1 -2 m - \mu, 1 -2 m - \mu  \ea   \right. \right)\times \cr
\fl & \times& G_{2,4}^{4,0}\left(x \lv
  \ba{l} 0, -1 + 2 m + \mu \\ -p, -p, -1 -p + 2 m + \mu, -1 -p + 2 m + \mu  \ea \right. \right).
\ena}
Then, the  integral of Meijer's G-functions products properties provides the partition function expression
\bea\label{part00}
 Z = \mbox{exp}\left[B_{m,\mu}\left(\frac{d}{dA}\right)^{2} \right]\sum_{n = 0}^{\infty} 
\left(e^{-A}\right)^n.
\ena
From (\ref{thermal03}),   using the result 
 \bea\label{dmatrixmv} 
\langle n+p| \rho^{(p)} |n+p\rangle  =   \frac{\mbox{exp}\left[B_{m,\mu}\left(\frac{d}{dA}\right)^{2} \right]\left(e^{-A}\right)^n}{Z}
\ena
and setting $\bar{n}_A =(e^{A} -1)^{-1}$, we get the following integration  equality
{\footnotesize
\bea 
\fl &&\frac{1}{\bar{n}_A+1}\left(\frac{\bar{n}_A}{\bar{n}_A+1}\right)^n  \frac{\Gamma(n+1)^2\Gamma(n+2m+\mu)^2}
{\Gamma(n+p+1)\Gamma(n+p+2m+\mu)}  =   \int_{0}^{\infty} dx \, x^{n+p} P(x) \times \cr
\fl  &&  G_{2,4}^{4,0}\left(x \lv  
  \ba{l}  0, -1 + 2 m + \mu \\ -p, -p, -1 -p + 2 m + \mu, -1 -p + 2 m + \mu  \ea \right. \right).\nonumber
\ena
}
After performing  the exponent change  $ n+p = s-1 $  of $x = |z|^2, $ in order to get  the Stieltjes moment problem, 
we arrive at the $P$-function  obtained as 
{\footnotesize
\bea\label{dmatrix08}
\fl & &P(|z|^2) =    \frac{1}{\bar{n}_A}\left(\frac{\bar{n}_A+1}{\bar{n}_A}\right)^p 
\frac{ G_{2,4}^{4,0}\left(\frac{\bar{n}_A+1}{\bar{n}_A} |z|^2 \lv  
  \ba{l}  0, -1 + 2 m + \mu \\ -p, -p, -1 -p + 2 m + \mu, -1 -p + 2 m + \mu  \ea \right. \right)}
{ G_{2,4}^{4,0}\left(|z|^2 \lv  
  \ba{l}  0, -1 + 2 m + \mu \\ -p, -p, -1 -p + 2 m + \mu, -1 -p + 2 m + \mu  \ea \right. \right)}
\ena
} 
which  obeys the normalization to unity condition (\ref{thermal04}).\\\\
Then, the diagonal representation of the normalized density operator in terms of the  GPAH-CS
projector (\ref{thermal03}) takes the form
 \bea 
\fl \rho^{(p)}  = \frac{1}{\bar{n}_A}\left(\frac{\bar{n}_A+1}{\bar{n}_A}\right)^p\int_{\IC} d^2 z \, \omega_p(|z|^2; m) 
|z;m\rangle_p \mathfrak S_{2,4}^{4,0}(|z|^2; {\bar{n}_A}) \, _{p}\langle z;m|     
\ena
with $ \mathfrak S_{2,4}^{4,0}(|z|^2, {\bar{n}_A})$ the Meijer's G-functions quotient given in (\ref{dmatrix08}).
\ni Using  the relations (\ref{eq96}), (\ref{eq97}), and the definition (\ref{obaverage00}),   the 
pseudo-thermal expectation values of the number operator and of its 
  square  coincide with  (\ref{numbtherm00}), where  $\bar{n}_o$ is replaced by  $\bar{n}_A$. 
  Therefore, the thermal intensity correlation function (\ref{thermcor}) and the thermal analogue of the Mandel parameter (\ref{thermcor00})
 have identical expressions as in (\ref{numbtherm01}) and (\ref{numbtherm03}), respectively, with $\bar{n}_A$ instead of $\bar{n}_o$.
\eni 
\subsubsection{Case where $f$ is not constant} \quad \\\\
Consider the linear function $\xi$ depending on the variable $r_{m,n}(k)\equiv r_{m+n-k+1}$:
\bea \label{eq100}
\xi(r_{m,n}(k); b; d) = b r_{m,n}(k) + d = 2b(m+n+\nu -k) + d 
\ena
 where we set $\mu = 2\nu$, and the  the product:
\bea  \label{eq101}
\prod_{k = 0}^{n - 1} \xi(r_{m,n}(k); b; d) = \prod_{k = 0}^{n - 1}(c(m+n+\nu -k + d/c))
\ena
where we put $c = 2 b$. After  a straightforward computation, we obtain: 
\bea  \label{eq102}
\prod_{k = 0}^{n - 1}\xi(r_{m,n}(k); c; d) = (-c)^n(-m-n-\nu -d/c)_n. 
\ena
Using the result $(-a)_k = (-1)^k \dis {a ! \over (a - k)!}$, we have:
\bea  \label{eq103}
\prod_{k = 0}^{n - 1}\xi(r_{m,n}(k); c; d) = (-c)^n{\Gamma(m+n+\nu +d/c + 1) \over \Gamma(m+\nu +d/c + 1)}. 
\ena
 {
We consider as  example of non constant function : 
$f(r_{m,n}(k)) = \sqrt{\xi(r_{m,n}(k); -1, 1)}.$}
We find
\bea  \label{eq104}
\prod_{k = 0}^{n - 1}f(r_{m,n}(k) = \sqrt{\Gamma(n+ m + \nu) \over \Gamma(m + \nu)},
\ena
and  the expansion coefficient reads: 
{\footnotesize
\bea  \label{eq105}
 K_n^p(m)
& = & \sqrt{{\Gamma(n+1)^2\Gamma(n+ 2 m + 2\nu)^2  \Gamma(m + \nu)
\over \Gamma(2 m + 2\nu)  \Gamma(n + p +1)  \Gamma(n + p + 2 m + 2\nu) \Gamma(n+ m + \nu)}}.\\
\ena}
\bei
\item[(i)]{\it Normalization}\\\\
The normalization factor (\ref{eq32}) becomes:
{\footnotesize
\bea \label{eq106}
\fl {\cal N}_p(|z|^2; m) & = & \left[{\Gamma(p+1)  \Gamma(2 m + p + 2\nu) \over  \Gamma(2 m + 2\nu)}\,
{}_3F_3\left(\ba{lcr} 1+p, 2m + p + 2\nu, m+\nu & ; & \\
 1, 2 m + 2\nu, 2 m + 2\nu & ; & |z|^2\ea \right)\right]^{-1/2},
 \ena}
where  $_3F_3$ is the generalized  hypergeometric function. In terms 
of Meijer's G-function
{\footnotesize
\bea \label{eq107}
\fl  {\cal N}_p(|z|^2; m) = \left[\Gamma(2 m + 2\nu)\, G_{3,4}^{1,3}\left(- {|z|^2 } \lv
 \ba{l}-p, 1 -2 m -p - 2\nu, 1 - m - \nu \\
 0  ,  0, 1 -2 m - 2\nu, 1 -2 m - \nu  \ea   \right. \right)\right]^{-1/2}
\ena}
For $p = 0$,
{
 \bea \label{eq108}
 \fl {\cal N}_0(|z|^2; m) &=  &[_1F_1(m + \nu; m + \nu; |z|^2)]^{-1/2}  \cr
 \fl   & = & [2^{2 m + 2\nu -1}\,|z|^{\half-\nu - m} e^{\half |z|^2}\, I_{ m + \nu - 1/2}(\half |z|^2) \Gamma( m + \nu +\half)]^{-1/2},
 \ena }
where $_1F_1$ is the confluent hypergeometric function, $I_\nu$-the modified Bessel function of the first kind.\\\\
The explicit form of the GPAH-CS corresponding  to the  Jacobi polynomials and hypergeometric functions, defined for any finite $|z|$, follows from (\ref{eq30}):
{\footnotesize
\bea \label{eq109}
\fl \lv z; m\rs_p & =  &{\cal N}_p(|z|^2; m) \, \times \cr 
\fl & & \times \sum_{n = 0}^\infty \sqrt{\dis { \Gamma(2 m + 2\nu)  \Gamma(n + p +1)  \Gamma(n + p + 2 m + 2\nu)\Gamma(n + m +\nu)
\over  \Gamma(n+ 2 m + 2\nu)^2 \Gamma(m +\nu)} }\ {z^n \over n!} \lv n + p\rs \pt
\ena}
\begin{figure}[htbp]
\begin{center}
\includegraphics[width=11cm]{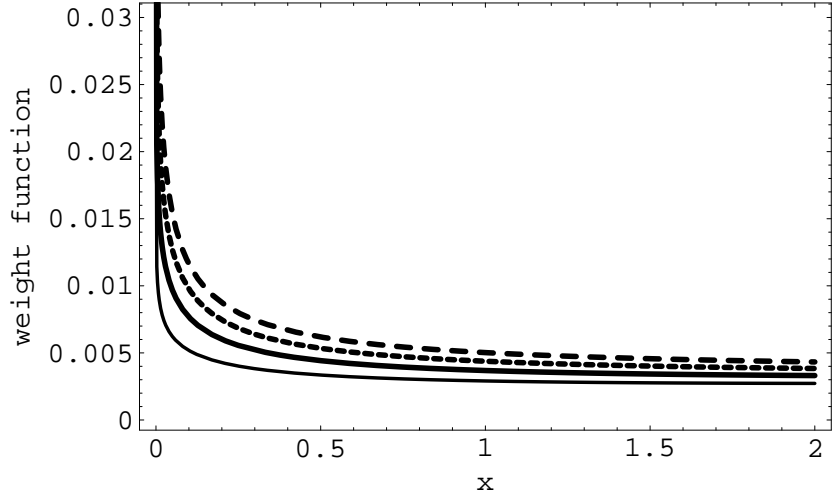}
\end{center}
\ni \caption[]{
Plots of the  weight function (\ref{eq113}) of  the
 GPAH-CS (\ref{eq109}) versus  $x = |z|^2$   with  the parameters $m = 3$, $\nu = 2.8$ and for different  values of  the photon added number $p = 1$ (thin solid  line), $p = 2$ (solid line), 
$p = 3$ (dot line), and $p = 4$ (dashed line). }
\end{figure}
\item[(ii)]{\it Overcompleteness}\\\\
\ni  The relation (\ref{eq39}) gives in this case 
{\footnotesize
\bea \label{eq110}
\fl  \int_0^\infty dx\ x^{n + p}\ g_p(x; m) & = &
{  \Gamma(m +\nu) \Gamma(n+1)^2 \Gamma(n+ 2 m + 2\nu)^2 \over 
\Gamma(2 m + 2\nu) \Gamma(n+p+1) \Gamma(n+ m +\nu) \Gamma(n+ p +2 m + 2\nu)}
\ena}
where the function $g_p$ is related to the measure $\omega_p$ by (\ref{eq61}).
Performing the variable change $n + m \to s - 1$ in   Eq. (\ref{eq110}) gives: 
{\footnotesize
\bea \label{eq111}
 \int_0^\infty dx \ x^{s-1}\ h_p(x; m) & = & \dis  {\Gamma(s-p)^2  \Gamma(s-p - 1 + 2 m + 2\nu)^2\over
 \Gamma(s) \Gamma(s-p - 1 +  m + \nu) \Gamma(s - 1 + 2 m + 2\nu)} 
\ena}
where $h_p(x; m) = g_p(x;m) \dis{ \Gamma(2 m + 2\nu) \over  \Gamma(m +\nu)}$. \ 
Application of the Mellin-inversion theorem (\ref{eq41}) provides:  
{\footnotesize
\bea \label{eq112}
  h_p(x; m) & = & G_{3,4}^{4,0}\left(x \lv
  \ba{l} 0, -1 + 2 m + 2\nu,-p-1+m+\nu \\ -p, -p, -1 -p + 2 m + 2\nu, -1 -p + 2 m + 2\nu  \ea \right. \right).
\ena}
{
Then, the weight function is given by 
{\small
\bea \label{eq113}
\fl \omega_p(|z|^2 ; m)& = & {\Gamma( m + \nu) \over \pi}\, G_{3,4}^{1,3}\left(- {|z|^2 } \lv
 \ba{l}-p, 1 -2 m -p - 2\nu, 1 - m - \nu \\
 0  ,  0, 1 -2 m - 2\nu, 1 -2 m - \nu  \ea   \right. \right)\, \times \cr
\fl & & \times G_{3,4}^{4,0}\left(|z|^2\lv
  \ba{l}  p, -1 + p + 2 m + 2\nu, -1+m+\nu  \\ 0, 0, -1  + 2 m + 2\nu, -1  + 2 m + 2\nu  \ea \right.\right)
\ena}
where we have used  (\ref{eq61}), (\ref{eq107}) and  the multiplication formula of the Meijer's G-function (\ref{eq64b}).
For $p = 0$, the function $h_p$ 
 is reduced to
\beano 
h_0(|z|^2; m) = G_{1,2}^{2,0}\left(x\lv
  \ba{l} -1+m+\nu  \\ 0, -1 + 2 m + 2\nu  \ea \right.\right)
\enano
and the weight function becomes  
{\small
\bea \label{eq114}
\fl  \omega_0(|z|^2; m) = {1\over \pi}\,
G_{1,2}^{1,1}\left(-|z|^2\lv
  \ba{l}1-m-\nu  \\ 0, 1-2 m- 2 \nu  \ea \right.\right).
 G_{1,2}^{2,0}\left(|z|^2\lv
  \ba{l} -1+m+\nu  \\ 0, -1 + 2 m + 2\nu  \ea \right.\right). \\ \nonumber
\ena}
In Figure 7, we plot the weight function (\ref{eq113}) versus $x = |z|^2$ for $m = 3$, $\nu = 2.8$ and different values of the photon-added number  $p = 1, 2, 3, 4$.  All the curves are positive, this confirm the positivity of the weight function for the parameter $\nu > 0$. As in the previous cases, the weight function presents a singularity at $x = 0$  and tends to zero for $x \to \infty.$}
\item[](iii){\it Photon number statistics}
\begin{figure}[htbp]
\begin{center}
\begin{minipage}{.45\textwidth}
\includegraphics[width=7cm]{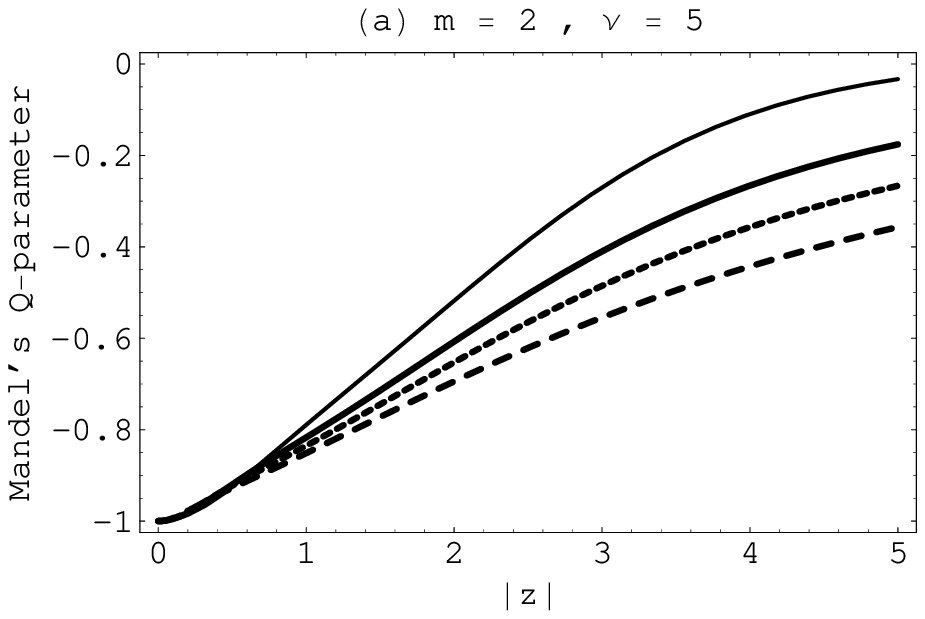}
\end{minipage} 
 \begin{minipage}{.45\textwidth}
\includegraphics[width=7cm]{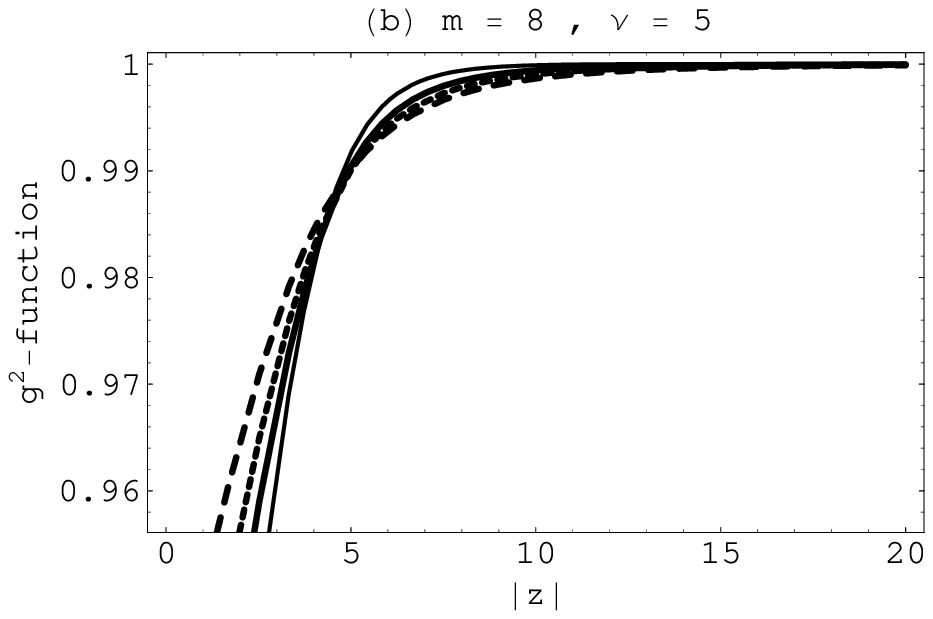}
\end{minipage}  
\end{center}
\ni \caption[]{
 Plots of: $(a)$ the Mandel Q-parameter  (\ref{eq117}) of the GPAH-CS (\ref{eq109}) versus   $|z|$ with the parameters 
$m = 2$ and $\nu = 5$ and for various values of the photon-added number $p$ with  $p= 1$ (thin solid  line),  $p = 3$ (solid line), $p = 5$ (dot line) and $p = 8$ (dashed line); 
 $(b)$ Second-order correlation function (\ref{eq118}) versus  $|z|$, with the parameters $m = 8$ and $\nu = 5$  for various values of the photon-added number $p$ with  $p= 1$ (thin solid  line), 
 $p = 3$ (solid line), $p = 5$ (dot line) and $p = 8$ (dashed line). } \label{Fig1_Jacobi2}
\end{figure}
\newline
\begin{figure}[htbp]
\begin{center}
\begin{minipage}{.45\textwidth}
\includegraphics[width=7cm]{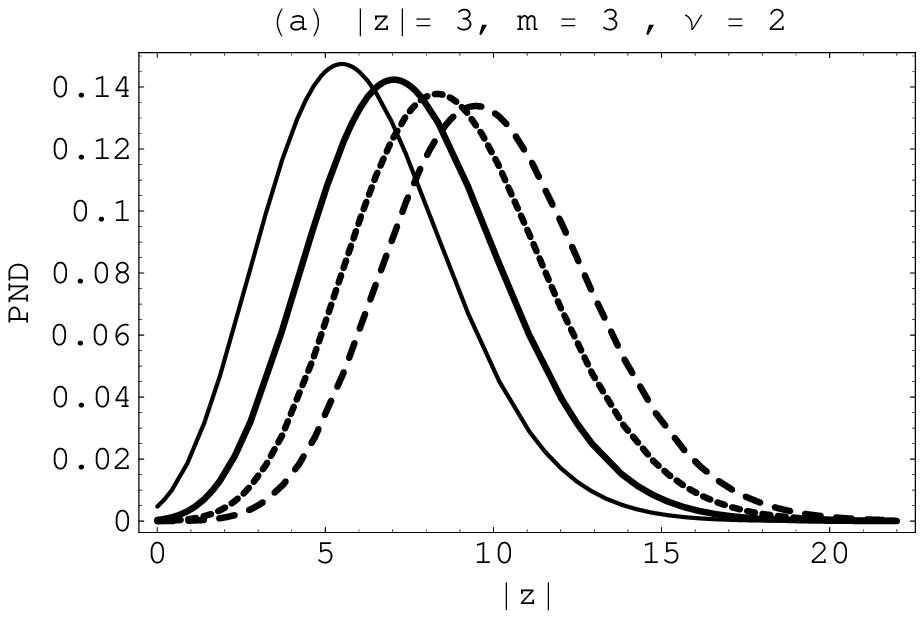}
\end{minipage} 
 \begin{minipage}{.45\textwidth}
\includegraphics[width=7cm]{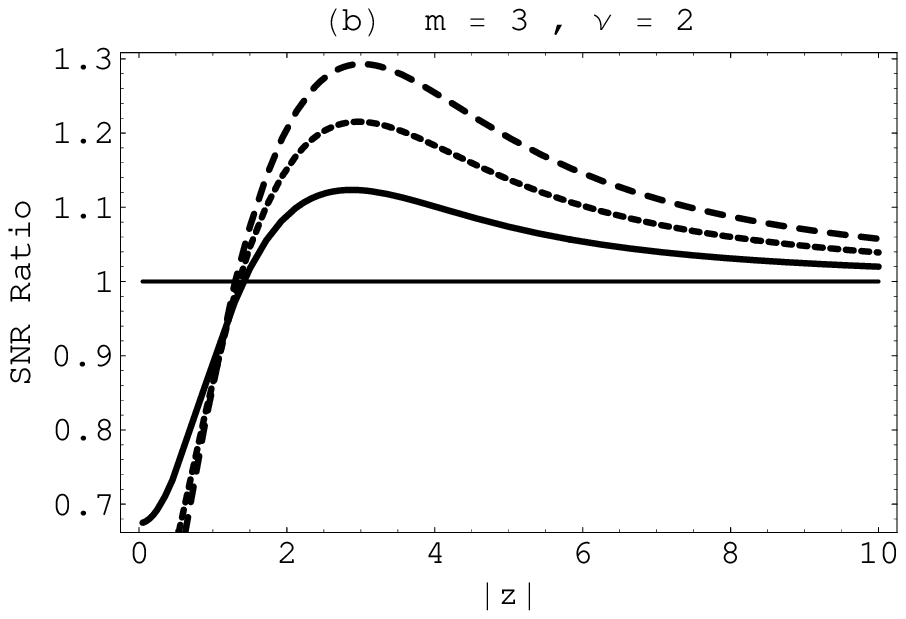}
\end{minipage}  
\end{center}
\ni \caption[]{
(a) Plot of the PND  (\ref{PND_Jacobi2}) of the GPAH-CS(\ref{eq109}) versus the photon number $n$, with $|z|=3, m = 3, \nu = 2$ and  for different values of the photon-added number $p$ with $p= 0$ (thin solid  line),
 $p = 1$ (solid line), $p = 2$ (dot line) and $p = 3$ (dashed line). \quad \\
 (b) 
Plot of the SNR (\ref{SNR_Jacobi2}) ratios of GPAH-CS (\ref{eq109})  to the corresponding GAH-CS versus  $|z|$, with $m = 3, \nu = 2$ and for 
various values  of the photon-added number $p$ with $p= 0$ (thin solid  line), $p = 1$ (solid line), $p = 2$ (dot line) and $p = 3$ (dashed line)} \label{Fig2_Jacobi2}
\end{figure}
\newline 
Using   the expressions 
 (\ref{eq105}) and (\ref{eq106}) of the factors $K_n^p(m)$ and ${\cal N}_p(|z|^2; m)$, respectively,   
we obtain  the expectation values (\ref{eq46}) and (\ref{eq47}) as:
{\footnotesize
\bea \label{eq115}
\fl \li N_m \rs &=  & (m + p)\  {_4F_4\left(\ba{lcr}1+p,m+p+1, p+2m+ 2\nu, m+\nu & ; & \\
         1,m+p, 2 m + 2\nu, 2 m + 2\nu &; &|z|^2\ea\right) \over
  _3F_3\left(\ba{lcr}1+p,2m +p +2\nu, m \nu & ; & \\ 1,  2 m + 2\nu, 2 m + 2\nu & ; & |z|^2\ea \right)} \\\label{eq116}
\fl \li N_m^2 \rs &= &  (m + p)^2\ 
 {_5F_5\left(\ba{lcl}1+p,m+\nu , m+p+1,m+p+1, p+2m+ 2\nu &;&\\
                                    1,m+p,m+p,  2 m +2 \nu, 2 m + 2\nu & ; & |z|^2 \ea \right) \over
  _3F_3\left(\ba{lcl}1+p,2m +p +2\nu, m \nu &;& \\
                1,  2 m + 2\nu, 2 m + 2\nu &;& |z|^2\ea \right)} \pt \ena }
Then, the Mandel Q-parameter is reduced to:
\bea \label{eq117}
Q = (m + p) \left[ {_5{\cal F}_5(|z|^2; m,p)\over  _4{\cal F}_4(|z|^2; m, p} -
     { _4{\cal F}_4(|z|^2; m, p)\over _3{\cal F}_3 (|z|^2; m, p)} \right] - 1.
\ena
The second order correlation  function (\ref{eq44}) follows as 
\bea  \label{eq118}
g^2 = {_3{\cal F}_3 (|z|^2; m, p)}\,{(m+p) _5{\cal F}_5(|z|^2; m, p) - _4{\cal F}_4(|z|^2; m, p)
 \over (m+p) _4{\cal F}_4(|z|^2; m, p)^2}
\ena
where $_3{\cal F}_3$, $_4{\cal F}_4$ and  $_5{\cal F}_5$ are the generalized hypergeometric functions:
{\footnotesize
\beano
\fl _3{\cal F}_3 (|z|^2; m, p) & = & _3F_3\left(\ba{lcl}1+p,2m +p +2\nu, m \nu &;&\\
                                      1,  2 m + 2\nu, 2 m + 2\nu &;& |z|^2\ea \right)\\
\fl  _4{\cal F}_4(|z|^2; m, p) & = & _4F_4\left(\ba{lcl}1+p,m+p+1, p+2m+ 2\nu, m+\nu  & ; & \\
                                 1,m+p, 2 m + 2\nu, 2 m + 2\nu &; &|z|^2\ea \right) \\
\fl _5{\cal F}_5(|z|^2; m, p) & = & _5F_5\left(\ba{lcl}1+p,m+\nu , m+p+1,m+p+1, p+2m+ 2\nu &;&\\
                                    1,m+p,m+p,  2 m +2 \nu, 2 m + 2\nu & ; & |z|^2 \ea \right) \pt
\enano}
The PND (\ref{PND}) reads as 
{\footnotesize
\bea\label{PND_Jacobi2}
 \fl {\cal P}_n^p(|z|^2; m) = {\Gamma(2m+2\nu)^2\Gamma(n+2m+2\nu) \Gamma(n+1) \Gamma(n-p +m+\nu)\over 
\Gamma(p+2m+2\nu)\Gamma(p+1) \Gamma(n - p +2m+2\nu)^2  \Gamma(m+\nu)\, _3{\cal F}_3 (|z|^2; m, p) }\, 
{|cz|^{2n} \over (n!)^2}.
\ena}
For $p = 0$, this reduces to PND of the conventional GAH-CS. 
\bea\label{PND_Jacobi2_p=0}
 \fl {\cal P}_n^p(|z|^2; m) = {\Gamma(2m+\mu) \Gamma(n+m+\nu)\over 
\Gamma(n+2m+2\nu)  \Gamma(m+\nu)\, _1F_1 (m+\nu; 2m + 2 \nu; |z|^2) }\, 
{|cz|^{2n} \over n!}.
\ena
\ni Finally,  the SNR  (\ref{SNR}) gives 
{\footnotesize
\bea \label{SNR_Jacobi2}
\fl \sigma^p(|z|^2; m) = {(m + p)\, _4{\cal F}_4(|z|^2; m, p) - p\, _3F_3(1+p ; 1; |cz|^2) \over 
(m+ p) \sqrt{_5{\cal F}_5(|z|^2; m,p) _3{\cal F}_3(|z|^2; m,p) - (_4{\cal F}_4(|z|^2; m,p))^2}}.\\
\ena}
 \item[(iv)]{\it  Thermal statistics}\\\\
 Starting with the  normalized density operator expression (\ref{dmatrix07}),  the $Q$-distribution or Husimi 
distribution analogue in the situation of the GPAH-CS (\ref{eq109}) 
 is provided as follows 
{\footnotesize
\bea
\fl _{p}\langle z;m| \rho^{(p)} |z;m\rangle_p= \frac{\mbox{exp}\left[B_{m,\mu}\left(\frac{d}{dA}\right)^{2} \right]}{Z}
\frac{ _3F_3\left(\ba{lcr} 1+p, 2m + p + 2\nu, m+\nu  & ; & \\
 1, 2 m + 2\nu, 2 m + 2\nu &  ; &  |z|^2 e^{-A}\ea\right)}
{ _3F_3\left(\ba{lcr}1+p, 2m + p + 2\nu, m+\nu & ; & \\
 1, 2 m + 2\nu, 2 m + 2\nu & ; & |z|^2 \ea \right)}\nonumber
\ena}
and in terms of Meijer's G functions,
{\footnotesize
\bea\label{dmatrix09}
 \fl &&_{p}\langle z;m| \rho^{(p)} |z;m\rangle_p =
 \frac{\mbox{exp}\left[B_{m,\mu}\left(\frac{d}{dA}\right)^{2} \right]}{Z}\,
\frac{G_{3,4}^{1,3}\left(- {|z|^2e^{-A}} \lv
 \ba{l}-p, 1 -2 m - 2\nu -p, 1-m-\nu  \\ 0 , 0, 1 -2 m - 2\nu, 1 -2 m - 2\nu  \ea   \right. \right)}
 {G_{3,4}^{1,3}\left(- {|z|^2 } \lv
 \ba{l}-p, 1 -2 m - 2\nu -p, 1-m-\nu  \\ 0 , 0, 1 -2 m - 2\nu, 1 -2 m - 2\nu  \ea   \right. \right)}.
\ena}
After an   angular integration, taking    $x = |z|^2, $ from the condition (\ref{thermal02}) we obtain 
{\footnotesize
\bea\label{dmatrix10}
 \fl\mbox{Tr}  \rho^{(p)} &=&\frac{\mbox{exp}\left[B_{m,\mu}\left(\frac{d}{dA}\right)^{2} \right]}{Z}\int_{0}^{\infty} dx \, x^p \, 
 G_{3,4}^{4,0}\left( {x } \lv
 \ba{l} 0, -1 + 2 m + 2\nu, -1-p + m+ \nu  \\ -p, -p, -1-p + 2 m + 2\nu, -1-p + 2 m + 2\nu   \ea   \right. \right)\cr
\fl & \times& G_{3,4}^{1,3}\left(- {x e^{-A}} \lv
 \ba{l}-p, 1 -2 m - 2\nu -p, 1-m-\nu  \\ 0 , 0, 1 -2 m - 2\nu, 1 -2 m - 2\nu  \ea   \right. \right).
\ena }
Applying the  integral of Meijer's G-functions products properties to (\ref{dmatrix10}) gives  the partition function expression
provided  in (\ref{part00}).\\
From the same diagonal elements in the states basis $|n+p\rangle $  as derived   in (\ref{dmatrixmv}),  the quasi-distribution function 
$P(|z|^2)$ is determined through the relation  
{\footnotesize
\bea 
\fl&&\frac{1}{\bar{n}_A+1}\left(\frac{\bar{n}_A}{\bar{n}_A+1}\right)^n  \frac{\Gamma(n+1)^2\Gamma(n+2m+\mu)^2}
{\Gamma(n+p+1)\Gamma(n+p+2m+\mu)} =  \int_{0}^{\infty} dx \, x^{n+p} P(x) \times \\
\fl && G_{3,4}^{4,0}\left( {x } \lv
 \ba{l} 0, -1 + 2 m + 2\nu, -1-p + m+ \nu  \\ -p, -p, -1-p + 2 m + 2\nu, -1-p + 2 m + 2\nu   \ea   \right. \right).\nonumber
\ena}
After performing  the exponent change  $ n+p = s-1 $  of $x = |z|^2, $ in order to get  the Stieltjes moment problem, 
we arrive at the $P$-function  obtained as 
{\footnotesize
\bea\label{dmatrix11}
\fl &P(|z|^2) = \frac{1}{\bar{n}_A}\left(\frac{\bar{n}_A+1}{\bar{n}_A}\right)^p 
\frac{ G_{3,4}^{4,0}\left( {\frac{\bar{n}_A+1}{\bar{n}_A} |z|^2} \lv
 \ba{l} 0, -1 + 2 m + 2\nu, -1-p + m+ \nu  \\ -p, -p, -1-p + 2 m + 2\nu, -1-p + 2 m + 2\nu   \ea   \right. \right)}
{ G_{3,4}^{4,0}\left( {|z|^2} \lv
 \ba{l} 0, -1 + 2 m + 2\nu, -1-p + m+ \nu  \\ -p, -p, -1-p + 2 m + 2\nu, -1-p + 2 m + 2\nu   \ea   \right. \right)}.
\ena}
satisfying  the normalization to unity condition (\ref{thermal04}).\\
Then, the diagonal representation of the normalized density operator in terms of the  GPAH-CS
projector (\ref{thermal03}),  takes the
form
 \bea 
\fl \rho^{(p)}  = \frac{1}{\bar{n}_A}\left(\frac{\bar{n}_A+1}{\bar{n}_A}\right)^p\int_{\IC} d^2 z \, \omega_p(|z|^2; m) 
|z;m\rangle_p \mathfrak S_{3,4}^{4,0}(|z|^2, {\bar{n}_A}) \, _{p}\langle z;m|     
\ena
with $ \mathfrak S_{3,4}^{4,0}(|z|^2, {\bar{n}_A})$ the Meijer's G-functions quotient given in (\ref{dmatrix11}).
\ni From the relations (\ref{eq115}), (\ref{eq116}),  and the definition (\ref{obaverage00}),  after computing  the 
pseudo-thermal expectation values of the number operator and of its 
  square, the thermal intensity correlation function (\ref{thermcor}) and the thermal analogue of the Mandel parameter (\ref{thermcor00}) are obtained as above, with their 
  expressions as in (\ref{numbtherm01}) and (\ref{numbtherm03}), respectively, with $\bar{n}_A$ instead of $\bar{n}_o$.
\eni
\subsection{Analysis of photon number statistics graphics}
\ni 
 In Figures 2, 5, 8,   we represent  the variations  of (a) the Mandel Q-parameter (\ref{eq72}, \ref{eq98}, \ref{eq117}),  
respectively, and (b) 
the second order correlation function  (\ref{eq73}, \ref{eq99}, \ref{eq118}), respectively,  
in terms of $|z|$,for different values of the
photon-added number $p$ and for fixed values of the derivative order parameter $m$,   and the polynomial parameter $\mu$ (or $\nu$). 
In these figures, the Mandel Q-parameter  is strictly  negative,     increases    
with the amplitude $|z|$,  and asymptotically tends to $0$. Besides, the second order correlation function is such that $0 < g^2 < 1$ 
and asymptotically tends to $1$.
Then,  the  GPAH-CS (\ref{eq56})  exhibit sub-Poisonnian distribution, and get close to Poissonian distribution as  $|z| \to \infty.$\\\\
 In Figures 3, 6, 9 (a),  the  PND is ploted  as a function of the photon number $n$,  
for a normal GAH-CS $p = 0$ along with a GPAH-CS for different values $p = 1, 2, 3$. 
  It follows that the shift in the  PND is more accentuated  as the number of photons  added increases.\\
In Figures 2, 4, 6 (b),  we plot  the ratio of the  SNR of various GPAH-CS, with  $p = 1, 2, 3$, to 
the SNR of a normal GAH-CS. The resulting curves reach their maximum over  $1$ and asymptotically tends to $1$, implying that 
the GPAH-CS    SNR is above the ordinary GAH-CS SNR. 
\section{Conclusion}
\ni In this paper, a set of non-classical states, i.e, the
generalized photon-added  associated 
hypergeometric coherent states 
(GPAH-CS), have been considered. 
These states are obtained by repeated applications of the raising operator $a_m^\dag$ on the generalized associated 
hypergeometric coherent states (GAH-CS). 
The required  Klauder  CS minimal set of conditions,
i.e, the label continuity, the 
normalizability and the  overcompleteness have been discussed.  This latter property,    evidenced through    moment problems   
explicitly solved by using the Mellin inversion theorem in terms of the Meijer's G-function,    has led  to reproducing
kernel analysis. Statistical quantities  like the Mandel Q-parameter and  the second-order correlation function have been derived
 from the number operator and its 
square expectation values. Besides, the signal-to-quantum-noise ratio (SNR), important for characterizing noiseless amplification of 
a coherent state, and also the photon 
number distribution (PND) evaluating the addition of photons for a given coherent light,  have been  analyzed. All these physical features 
have been   depicted for appropriate values and discussed. 
In order to examine the thermal statistics  of these states, 
the density matrix of a quantum canonical ideal  gas of the 
 system in thermodynamic equilibrium  has been first given. Then,    the Q-distribution
or Husimi distribution, the diagonal representation of the density operator 
 and the P-distribution function  expressions in the GPAH-CS have been obtained. 
As interesting 
applications,  the Hermite, Laguerre, Jacobi polynomials,  and the  hypergeometric functions have been studied. Their related classes of 
GPAH-CS as well the GAH-CS have been built, showing sub-Poissonian distribution, and their thermal statistics provided.
 For instance  
as $|z|$ takes  large values , the Mandel parameter and the
second-order correlation function tend  to $0$ and $1$, respectively. Thus,    the GPAH-CS become classical for
large values of $|z|$. Moreover, increasing the number of photons leads to a shift of the PND.
 
\end{document}